\documentclass[onecolumn,nameyear,12pt]{autart}
\makeatletter
\def\@biblabel#1{}
\makeatother
\usepackage{float}
\usepackage{ifpdf}
\usepackage{graphicx,amsmath,amssymb,mathrsfs,lineno,cite}
\usepackage{epstopdf}

\begin{document}

\begin{frontmatter}

\title{Analysis of Quantum Linear Systems' Response to Multi-photon States}

\author[zhang]{Guofeng Zhang}\ead{Guofeng.Zhang@polyu.edu.hk}

\address[zhang]{Department of Applied Mathematics, The Hong Kong Polytechnic University, Hong Kong}


\begin{abstract}
The purpose of this paper is to present a mathematical framework for analyzing the response of quantum linear systems driven by multi-photon states. Both the factorizable (namely, no correlation among the photons in the channel) and unfactorizable multi-photon states are treated. Pulse information of multi-photon input state is encoded in terms of tensor, and response of quantum linear systems to multi-photon input states is characterized by tensor operations. Analytic forms of output correlation functions and output states are derived. The proposed framework is applicable no matter whether the underlying quantum dynamic system is passive or active.  The results presented here generalize those in the single-photon setting studied in (\cite{Milburn08}) and (\cite{ZJ13}). Moreover, interesting multi-photon interference phenomena studied in (\cite{SRZ06}), (\cite{Ou07}), and (\cite{BDS+12}) can be reproduced in the proposed framework.
\end{abstract}

\end{frontmatter}

\section{Introduction} \label{sec:introduction}

Analysis of system response to various types of input signals is fundamental to control systems engineering. Step response enables a control engineer to visualize system transient behavior such as rise time, overshoot and settling time; frequency response design methods are among the most powerful methods in classical control theory; response analysis of linear systems initialized in Gaussian states driven by Gaussian input signals is the basis of Kalman filtering and linear quadratic Gaussian (LQG) control (see, e.g., \cite{AM71}; \cite{KS72}; \cite{AM79}; \cite{ZDG96}; \cite{QZ09}).

Over the last two decades, there has been rapid advance in experimental demonstration and theoretical investigation of quantum (namely, non-classical) control systems due to their promising applications in a wide range of areas such as quantum communication, quantum computation, quantum metrology, laser-induced chemical reaction, and nano electronics (\cite{GZ00}; \cite{RL00}; \cite{NC00}; \cite{DDA07}; \cite{WM08}; \cite{WM10};  \cite{VPB83}; \cite{HTC83}; \cite{YD84}; \cite{Gar93}; \cite{DJ99}; \cite{KBG01}; \cite{AA03}; \cite{YK03}; \cite{SvHM04}; \cite{MK05}; \cite{vHSM05}; \cite{CA07}; \cite{MvH07}; \cite{JNP08}; \cite{PR08}; \cite{BCS09}; \cite{GJ09}; \cite{LK09}; \cite{MR09}; \cite{NJD09}; \cite{YB09}; \cite{BBR10}; \cite{BT10}; \cite{BCR10}; \cite{DP10}; \cite{GJN10}; \cite{MNM10}; \cite{WS10}; \cite{MP11}; \cite{ZJ11}; \cite{AT12}; \cite{AMR12}; \cite{ZWL+12}; \cite{BQ13}). Within this program quantum linear systems play a prominent role. Quantum linear systems are characterized by linear quantum stochastic differential equations (linear QSDEs). In quantum optics, linear systems are widely used because they are easy to manipulate and, more importantly, linear dynamics often serve well as good approximation of more general dynamics (\cite{GZ00}; \cite{RL00}; \cite{WM08}; \cite{WM10}). Besides their broad applications in quantum optics, linear systems have also found applications in many other quantum-mechanical systems such as opto-mechanical systems (\cite[Eqs. (15)-(18)]{MHP+11}), circuit quantum electrodynamics (circuit QED) systems (\cite[Eqs. (18)-(21)]{MJP+11}), atomic ensembles (\cite[Eqs. (A1),(A4)]{SvHM04}), quantum memory (\cite[Eqs. (12)-13]{HCH+13}). From a signals and systems point of view, quantum linear systems driven by Gaussian input states have been studied extensively, and results like quantum filtering and measurement-based feedback control have been well established (\cite{WM10}).

In addition to Gaussian states there are other types of non-classical states, for example single-photon states and multi-photon states. Such states describe electromagnetic fields with a definite number of photons. Due to their highly non-classical nature and recent hardware advance, there is rapidly growing interest in the generation and engineering (e.g., pulse shaping) of photon states, and it is generally perceived that these photon states hold promising applications in quantum communication, quantum computing, quantum metrology and quantum simulations (\cite{CMR09}; \cite{GEP+98}; \cite{SRZ06}; \cite{Ou07}; \cite{BDS+12}; \cite{Milburn08}; \cite{GJN13}; \cite{HCH+13}). Thus, a new and important problem in the field of quantum control engineering is: How to characterize and engineer interaction between quantum linear systems and photon states? The interaction of quantum linear systems with continuous-mode photon states has recently been studied in the literature, primarily in the physics community. For example, interference phenomena of photons passing through beamsplitters have been studied, see, e.g., \cite{SRZ06}; \cite{Ou07}; \cite{BDS+12}. Milburn discussed how to use an optical cavity to manipulate the pulse shape of a single-photon light field (\cite{Milburn08}). Quantum filtering for systems driven by single-photon fields has been investigated in \cite{GJN13}, based on which nonlinear phase shift of coherent signal induced by single-photon field has been studied in \cite{CHJ12}. Intensities of output fields of quantum systems driven by continuous-mode multi-photon light fields have been studied in \cite{BCB+12}. In \cite{ZJ13} the response of quantum linear systems to single-photon states has been studied. Formulas for intensities of output fields have been derived. In particular, a new class of optical states, photon-Gaussian states, has been proposed.

In the analysis of the response of quantum linear systems to single-photon states, matrix presentation is sufficient because two indices are adequate: one for input channels, and the other for output channels. However, this is not the case in the multi-photon setting. In addition to indices for input and output channels, we need another index to count photon numbers in channels. As a result, tensor representation and operation are essential in the multi-photon setting. To be specific, multi-photon state processing by quantum linear systems can be mathematically represented in terms of tensor processing by transfer functions. The key ingredient for such an operation is the following (for the passive case). Let $E(t)=(E^{jk}(t))\in\mathbb{C}^{m\times m}$ be the transfer function of a quantum linear passive system with $m$ input channels. For each $j=1,\ldots,m$, let $\mathscr{V}_j(t_1,\ldots,t_{\ell_j})$ be an $\ell_j$-way $m$-dimensional tensor function that encodes the pulse information of the $j$-th  input channel containing $\ell_j$ photons. Denote the entries of $\mathscr{V}_j(t_1,\ldots,t_{\ell_j})$ by $\mathscr{V}_{j,k_1,\ldots,k_{\ell_j}}(t_1,\ldots,t_{\ell_j})$.  For all given $1\leq r_1,\ldots,r_{\ell_j}\leq m$, define an  $\ell_j$-way $m$-dimensional tensor $\mathscr{W}_j$ with entries given by the following multiple convolution
\begin{align*}
   & \mathscr{W}_{j,r_1,\ldots,r_{\ell_j}}(t_1,\ldots,t_{\ell_j}) \\
  =\sum_{k_1,\ldots,k_{\ell_j}=1}^m & \int_{-\infty}^\infty \cdots \int_{-\infty}^\infty E^{r_1k_1}(t_1-\iota_1)\cdots E^{r_{\ell_j}k_{\ell_j}}(t_{\ell_j}-\iota_{\ell_j})\mathscr{V}_{j,k_1,\ldots,k_{\ell_j}}(\iota_1,\ldots,\iota_{\ell_j})
 d\iota_1\ldots d_{\ell_j}  .\nonumber
\end{align*}
 It turns out that the tensors $\mathscr{W}_j$ ($j=1,\ldots,m$) encode the pulse information of the output field. That is, an $\ell_j$-way input tensor is mapped to an $\ell_j$-way output tensor by the quantum linear passive system.

The contributions of this paper are three-fold. First, the analytic form of the steady-state output state of a quantum linear system driven by a multi-photon input state is derived. When the quantum linear system is a beamsplitter (a static passive device), interesting multi-photon interference phenomena studied in (\cite{SRZ06}), (\cite{Ou07}), and (\cite{BDS+12}) are re-produced by means of our approach, see Examples 1,2,3. Second, when the underlying quantum linear system is not passive (e.g., a degenerate parametric amplifier), the steady-state output state with respect to a multi-photon input state is not a multi-photon state. In terms of tensor representation, a more general class of states is defined. Such rigorous mathematical description paves the way for multi-photon state engineering. Third, both the factorizable and unfactorizable multi-photon states are treated in this paper. Here a factorizable multi-photon state is a state for which the photons in a given channel are not correlated, while for an unfactorizable multi-photon state there exists correlation among the photons. This difference cannot occur in the single-photon state case.  Thus, the mathematical framework presented here is more general.

The rest of the paper is organized as follows.  Preliminary results are presented in Section \ref{sec:preliminaries}. Specifically,  quantum linear systems are briefly reviewed in Subsection \ref{sec:systems} with focus on stable inversion and covariance function transfer, in Subsection \ref{sec:tensor} several types of tensors and their associated operations are introduced.  The multi-photon state processing when input states are factorizable in terms of pulse shapes is studied in Section \ref{sec:factorable}. (Here the word ``factorizable'' means there is no correlation among photons in each specific channel.) Specifically, single-channel and multi-channel multi-photon states are presented in Subsections \ref{subsec:single-channel} and \ref{sec:multi_photon_def} respectively,  covariance functions and intensities of output fields are studied in Subsection \ref{subsec:covariance_intensity}, while an analytic form of steady-state output states is derived in Subsection \ref{sec:output_state}. The  unfactorizable case is investigated in Section \ref{sec:unfactorable}. Specifically,  unfactorizable multi-channel multi-photon states are defined in Sebsection \ref{subsec:general_photon}, the analytic form of the steady-state output state is presented in Subsection \ref{subsec:general_passive} where the underlying system is passive, the active case is studied in Subsection \ref{sec:active}. Some concluding remarks are given in  Section \ref{sec:conclusion}.

\emph{Notations.} $m$ is the number of input channels, and $n$ is the number of degrees of freedom of a
given quantum linear stochastic system. $|\phi\rangle$ denotes the
initial state of the system which is always assumed to be vacuum, $|0\rangle$ denotes the vacuum state of free fields.
Given a column vector of complex numbers or operators
$x
=
[
\begin{array}{ccc}
x_1 & \cdots & x_k%
\end{array}
]^T
$
where $k$ is a positive integer, define
$x^\#
=
[
\begin{array}{ccc}
x_1^\ast & \cdots & x_k^\ast%
\end{array}
]^T
$,
where the asterisk $\ast$ indicates complex conjugation or Hilbert
space adjoint. Denote $x^\dag = (x^\#)^T$. Furthermore, define the
doubled-up column vector to be
$
\breve{x}
=
[
\begin{array}{cc}
x^T & (x^\#)^T%
\end{array}
]^T
$.  Let $I_k$ be an
identity matrix and $0_k$ a zero square matrix, both of dimension $k$.
Define $J_k=\mathrm{diag}(I_k,-I_k)$ and
$
\Theta_k
=
[
\begin{array}{cccc}
0 & I_k; & -I_k & 0%
\end{array}%
]
$ (The subscript ``$k$'' is often omitted.) Then for
a matrix $X\in\mathbb{C}^{2j\times 2k}$, define $X^\flat:=J_k X^\dag J_j$. $\otimes_c$ denotes the Kronecker product. Given a function $f(t)$
in the time domain, define its two-sided Laplace transform (\cite[(13)]{Sog10}) to be
$F[s] = \mathscr{L}_b \{f(t)\}(s) := \int_{-\infty}^\infty e^{-st} f(t) dt $. Given two constant matrices $U$, $V\in \mathbb{C}^{r\times k}$, define $\Delta(U,V) = [ U ~ V; V^\# ~ U^\#]$.
Similarly, given time-domain matrix functions $E^-(t)$ and $E^+(t)$ of
compatible dimensions, define $\Delta(E^-(t),E^+(t)) = [E^-(t) ~ E^+(t);E^+(t)^\# ~  E^-(t)^\#]$. Given two operators $A$ and $B$, their commutator is defined to be $[A,B]:=AB-BA$. For any integer $r>1$, we write $\int_r$ for integration in the space $\mathbb{R}^r$. We also write $dt_{1\to r}$ for $dt_1\cdots dt_r$. Finally, given a column vector $a$, we use $a_j$ to denote its entries. Given a matrix $A$, we use $A^{jk}$ to denote its entries. Given a 3-way tensor $\mathcal{A}$ (also called a tensor of order 3), we use $\mathcal{A}_{ijk}$ to denote its entries; we do the similar thing for higher order tensors.

\section{Quantum linear systems and tensors}\label{sec:preliminaries}
This section records preliminary results necessary for the development of the paper. Quantum linear systems are briefly discussed is Subsection \ref{sec:systems}.  Tensors and their associated operations, the appropriate mathematical language to describe the interaction of a quantum linear system with multi-photon channels, is introduced in Subsection \ref{sec:tensor}.

\subsection{Quantum linear systems}\label{sec:systems}

In this subsection quantum linear systems are described in the input-output language, which makes it natural to present transfer of covariance function of input fields. Moreover, the input-output framework also enables the definition of the stable inversion of quantum linear systems.

\begin{figure}[ptb]
\centering
\includegraphics[width=2.0in]{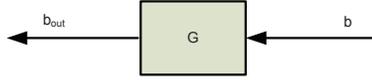}\caption{Quantum linear system $G$ with input $b$ and output $b_{\rm out}$}
\label{system}%
\end{figure}

\subsubsection{Fields and systems} \label{subsubsec:field_system}
In this part we set up the model which is a quantum linear system driven by boson fields, (\cite{GZ00}; \cite{WM08}; \cite{WM10}).

The triple $(S,L,H)$ provides a compact way for the description of open quantum systems (\cite{GJ09}; \cite{GJN10}; \cite{ZJ12}). Here the self-adjoint operator $H$ is the initial system Hamiltonian, $S$ is a unitary scattering operator, and $L$ is a coupling operator that describes how the system is coupled to its environment. The environment is an $m$-channel electromagnetic field in free space, represented by a column vector of annihilation operators $b(t)=[b_1(t),\cdots,b_m(t)]^T$. Let $t_0$ be the initial time, namely, the time when the quantum system starts interacting with its environment. Define a gauge process $\Lambda(t)$ by
$\Lambda(t)=\int_{t_0}^t b^\#(\tau)b^T(\tau)d\tau = \left(\Lambda^{jk}(t)\right)_{j,k=1,\ldots,m}$
with operator entries $\Lambda^{jk}(t)$ on the Fock space $\mathfrak{F}$ for
the free field (\cite{GZ00,WM08}). In this paper it is assumed that these quantum stochastic
processes are \emph{canonical}, that is, they have the following non-zero Ito products
\begin{align*}
dB_j(t)dB_k^\ast(t)=  &  ~ \delta_{jk}dt, ~ d\Lambda^{jk}dB_l^\ast(t)=\delta_{kl}dB_j^\ast(t),\label{Eq:CCR2}\\
dB_j(t)d\Lambda^{kl}(t)=  & ~  \delta_{jk}dB_l(t), ~ d\Lambda^{jk}(t)d\Lambda^{lm}(t)=\delta_{kl}d\Lambda^{jm}(t),~j,k,l=1,\ldots,m,\nonumber
\end{align*}
where $B(t)=[B_1(t),\cdots ,B_m(t)]^T$ is a column vector of the integrated field operators defined via $B(t):=\int_{t_0}^tb(r)dr$.
In the \emph{interaction picture} the stochastic Schrodinger's equation for the open quantum system driven by the free field $b(t)$ is, in Ito form (\cite[Chapter 11]{GZ00}),
\begin{equation} \label{eq:U}
dU(t,t_0) =
\left\{\mathrm{Tr}[(S -I_m)d\Lambda(t)^T]+dB^\dag(t)L-L^\dag SdB(t)-(\frac{1}{2}L^\dag L+iH)dt\right\}
U(t,t_0),  ~~  t\geq t_0,
\end{equation}
with $U(t,t_0)=I$ being an identity operator for all $t\leq t_0$.

Specific to the linear case, the open quantum linear system $G$ shown in
Fig.~\ref{system} represents a collection of $n$ interacting quantum harmonic
oscillators $a(t)=[a_1(t),\ldots,a_n(t)]^T$ (defined on a Hilbert space
$\mathfrak{H}_G$) coupled to $m $ boson fields $b(t)$ described above (\cite{GZ00,WM10,ZJ11,ZJ12}).
Here, $a_j$ ($j=1,\ldots,n$) is the annihilation operator of the $j$th oscillator satisfying the canonical commutation relations
$[a_j,a_k^\ast]=\delta_{jk}$. Denote $\breve{a}(t_{0})=\breve{a}$. The
vector operator $L\in\mathfrak{H}_G$ is defined as $L=C_-a+C_+a^\#$ with $C_-, C_+\in\mathbb{C}^{m\times n}$. The initial Hamiltonian $H\in\mathfrak{H}_G$ is
$H=\frac{1}{2}\breve{a}^\dag\Delta\left(\Omega_-,\Omega_+\right)\breve{a}$
with $\Omega_-,\Omega_+\in \mathbb{C}^{n\times n}$ satisfying
$\Omega_- = \Omega_-^\dag$ and $\Omega_+=\Omega_+^T$. By (\ref{eq:U}), the dynamic
model for the system $G$ is
\begin{eqnarray}
\dot{\breve{a}}(t)  &  = & A\breve{a}(t)+B\breve{b}(t),\ \
\breve{a}(t_0) = \breve{a}, \label{system-a}\\
\breve{b}_{\rm out}(t)   &  = & C\breve{a}(t)+D\breve{b}(t)  , 
\label{system-out}%
\end{eqnarray}
in which system matrices are given in terms of the physical parameters $S,L,H$, specifically,
\[
D = \Delta(S,0), \  C=\Delta(C_-,C_+), \ B=-C^\flat \Delta(S,0), \ A=-\frac{1}{2}C^\flat C-iJ_n\Delta(\Omega_-,\Omega_+).
\]
The \emph{transfer function} (impulse response function) for the system $G$ is
\begin{equation}
g_G(t)
:=
\left\{
\begin{array}{ll}%
\delta(t)D +  Ce^{At}B, & t\geq0,\\
0, & t<0.
\end{array}
\right.
\label{eq:gg}
\end{equation}
This, together with (\ref{system-a}) and (\ref{system-out}), yields
\begin{equation}
\breve{b}_{\rm out}(t)=Ce^{A(t-t_{0})}\breve{a}+\int_{t_0}^t g_G(t-r)\breve{b}(r)dr.
\label{eq:out_tf}%
\end{equation}

The system $G$ is said to be \emph{passive} if both $C^+=0$ and $\Omega^+=0$. The system $G$ is said to be \textit{asymptotically stable} if the matrix $A$ is Hurwitz (\cite[Section III-A]{ZJ11}).

Define matrix functions
\begin{align*}
g_{G^-}(t)
&  :=
\left\{
\begin{array}{ll}%
\delta(t)S -[%
\begin{array}{cc}%
C_- & C_+%
\end{array}
]e^{At}\left[
\begin{array}{c}%
C_-^\dag\\
-C_+^\dag%
\end{array}
\right]  S, & t\geq0,\\
0, & t<0,
\end{array}
\right.  \nonumber\\
\ g_{G^+}(t)
&  :=
 \left\{
\begin{array}{ll}%
-[%
\begin{array}{cc}%
C_- & C_+%
\end{array}
]e^{At}\left[
\begin{array}{c}%
-C_+^T\\
C_-^T%
\end{array}
\right]  S^\#, & t\geq0,\\
0, & t<0.
\end{array}
\right.   \label{eq:io}%
\end{align*}
(Note that when $G$ is passive, $g_{G^+}(t)\equiv0$.) With these functions, the transfer function in  (\ref{eq:gg}) can be re-written as
\begin{equation*}
g_G(t)=\Delta\left(g_{G^-}(t),g_{G^+}(t)\right)  . \label{eq:impulse}%
\end{equation*}

Finally, assume that the system (\ref{system-a})-(\ref{system-out}) is asymptotically stable. Letting
$t_0\rightarrow-\infty$ and noticing (\ref{eq:gg}), equation (\ref{eq:out_tf}) becomes
\begin{equation}
\breve{b}_{\rm out}(t)=\int_{-\infty}^\infty g_G(t-r)\breve{b}(r)dr ,
\label{eq:out_tf3}%
\end{equation}
which characterizes the steady-state relation between the input and the output.

\subsubsection{Stable inversion}
In this par some results for the stable inversion of quantum linear systems $G$ are recorded, which are used in the derivation of output states of quantum linear systems driven by multi-photon states, cf. Sections \ref{sec:factorable} and \ref{sec:unfactorable}.

For the transfer function $g_{G}(t)$ defined in (\ref{eq:gg}), let $\Xi_{G}[s]$ be its two-sided Laplace transform (see the \emph{Notations} part and  \cite[Eq. (13)]{Sog10}. Define a matrix function $g_{G^{-1}}(t)$ to be
\begin{equation}
g_{G^{-1}}(t):=\mathscr{L}_b^{-1}\{\Xi_{G}[s]^{-1}\}(t).
\label{eq:stable-inverse}%
\end{equation}
The following result is proved in \cite[Lemma 1]{ZJ13}.

\begin{lem}\label{lem:G_inv}
Assume that the system $G$ is asymptotically stable. Then
\begin{eqnarray}
g_{G^{-1}}(t) &=& \Delta\left(g_{G^-}(-t)^\dag, -g_{G+}(-t)^T\right)  .
\end{eqnarray}
\end{lem}

\emph{Remark 1.} Because the system $G$ is asymptotically stable, it has no zeros on the imaginary axis, $\Xi_{G}[s]^{-1}$ is well defined. It is worth pointing out that the matrix function $g_{G^{-1}}(-t)$ turns out to be the transfer function of the inverse system defined in \cite[(71)]{GJN10}.

\begin{lem}\label{lem:G_inv_2}
Assume the quantum linear system $G$ is asymptotically stable. Define an operator
\begin{equation*}\label{eq:b^-}
\breve{b}^-(t,t_0):=U(t,t_0)\breve{b}(t)U(t,t_0)^\ast , \ \ \ t\geq t_0.
\end{equation*}
Then
\begin{equation}\label{eq:b^-2}
\breve{b}^-(t,-\infty)
=
\int_{-\infty}^\infty g_{G^{-1}}(t-r) \breve{b}(r) dr.
\end{equation}
\end{lem}

\textbf{Proof.}~ Because $\breve{b}_{\rm out}(t)=U(t,t_0)^\ast\breve{b}(t)U(t,t_0)$, equation (\ref{eq:out_tf}) gives
\[
U(t,t_0)^\ast \breve{b}(t)  U(t,t_0) = Ce^{A(t-t_{0})}\breve{a}+\int_{t_0}^t g_G(t-r)\breve{b}(r)dr .
\]
That is,
\[
\breve{b}(t) =  Ce^{A(t-t_{0})}U(t,t_0)\breve{a}U(t,t_0)^\ast+\int_{t_0}^t g_G(t-r)\breve{b}^-(t,t_0) dr .
\]
Letting $t_0\to -\infty$ and noticing (\ref{eq:gg}) we have
\begin{equation}\label{eq:inverse_temp}
\breve{b}(t)
=
\int_{-\infty}^\infty g_G(t-r)\breve{b}^-(r,-\infty)dr.
\end{equation}
However, by Eq. (\ref{eq:stable-inverse}), we have $g_{G^{-1}} \ast g_G (t)   = \delta(t)$. 
Substituting it into (\ref{eq:inverse_temp}) yields (\ref{eq:b^-2}).

\emph{Remark 2.} Operators $\breve{b}^-(r,t_0)$ and $\breve{b}^-(r,-\infty)$ are formally defined mathematically, they may not correspond to physical variables. However, they do help in the derivation of the steady-state state of output fields.

\subsubsection{Steady-state covariance transfer}

Here we record esults concerning covariance function transfer by the quantum linear system $G$ defined in (\ref{system-a})-(\ref{system-out}). 

Assume the quantum linear system $G$ is in the vacuum state $|\phi\rangle$. Assume further that the input field is in a zero-mean state $\rho_{\rm f}$. (specific types of $\rho_{\rm f}$ will be studied in the sequel.) Denote the covariance functions of the input filed $b(t)$ and the output field $b_{\rm out}(t)$ by $R(t,r)$ and $R_{\rm out}(t,r)$ respectively, that is,
\begin{equation}\label{eq:R}
R(t,r)=\mathrm{Tr}[\rho_{\rm f}\breve{b}(t)\breve{b}^\dag(r)], \ \ R_{\rm out}(t,r)=\mathrm{Tr}[|\phi\rangle\langle\phi|\otimes\rho_{\rm f}\breve{b}_{\rm out}(t)\breve{b}^\dag_{\rm out}(r)].
\end{equation}
According to (\ref{eq:out_tf3}) and (\ref{eq:R}) we have
\begin{lem} \label{lem:covariance-transfer}
Assume that the system (\ref{system-a})-(\ref{system-out}) is asymptotically stable. Let the input field have covariance $R(t,r)$ defined in (\ref{eq:R}). The steady-state (namely $t_0\to-\infty$) output covariance function $R_{\rm out}(t,r)$ is
\begin{equation}
R_{\rm out}(t,r)
=
\int_{-\infty}^\infty\int_{-\infty}^\infty g_G(t-\tau_1)R(\tau_1,\tau_2)g_G(r-\tau_2)^\dag
d\tau_1 d\tau_2 .
\label{Eq:out_in}
\end{equation}
\end{lem}

In the frequency domain, we have
\begin{thm}\label{thm:spectral-transfer}
Assume that the system (\ref{system-a})-(\ref{system-out}) is asymptotically stable. If the input field is stationary with spectral density matrix $R[i\omega]$ (namely, the Fourier transform of $R(t,t)$), the output spectral density matrix is given by
\begin{equation}
R_{\rm out}[i\omega]
=
\Xi_G [i\omega]R[i\omega]\Xi_G [i\omega]^\dag .
\label{eq:R-transfer}%
\end{equation}
In particular, if the input field is in the vacuum state $|0\rangle$, that is, $R[i\omega] =
\left[
\begin{array}{cc}%
I_m & 0\\
0 & 0_m
\end{array}
\right]$, then the output state is a Gaussian state with output spectral density matrix
\begin{equation}
R_{\rm out}[i\omega]
=
\Xi_G [i\omega]\left[
\begin{array}{cc}%
I_m & 0\\
0 & 0_m
\end{array}
\right]\Xi_G [i\omega]^\dag .
\label{eq:R-transfer_2}%
\end{equation}
\end{thm}

In what follows we focus on the Gaussian input field states. Denote the initial joint system-field Gaussian state by $\rho_{0{\rm g}}=|\phi\rangle\langle\phi|\otimes\rho_{\rm f}$ where $\rho_{\rm f}$ is a Gaussian field state. Define the steady-state joint state
\begin{equation}
\rho_{\infty {\rm g}}:=\lim_{\begin{subarray}{c}t_0\rightarrow-\infty \\ t\rightarrow\infty\end{subarray}}U(t,t_0)\rho_{0{\rm g}}U(t,t_0)^\ast,
\label{eq:rho_inf_g}%
\end{equation}
and the steady-state output field state
\begin{equation*}
\rho_{\rm field, g}:=\mathrm{Tr}_{\rm sys}[\rho_{\infty {\rm g}}], \label{eq:rho_field}%
\end{equation*}
where the subscript ``sys'' indicates that the trace operation is performed with respect to the system. Moreover, if the input state $\rho_{\rm f}$ is stationary with spectral density $R[i\omega]$, according to Theorem \ref{thm:spectral-transfer}, $\rho_{\rm field, g}$ is the steady-state output field density with covariance function $R_{\rm out}[i\omega]$ given in (\ref{eq:R-transfer}). Finally, if $\rho_{\rm f}=|0\rangle\langle0|$, then $\rho_{\rm field, g}$ is stationary zero-mean Gaussian with $R_{\rm out}[i\omega]$ given in (\ref{eq:R-transfer_2}).

\emph{Remark 3.} Because  $\rho_{field,g}$  is obtained by tracing out the system, it is in general a mixed state. Moreover, $\rho_{\infty g}$ is in general not the vacuum state even if $\rho_f=|0\rangle\langle0|$. However,
if the system $G$ is \emph{passive}, then by (\ref{eq:R-transfer_2}),
\begin{equation}\label{R_out_vac}
R_{out} [i\omega]
=
\left[
\begin{array}{cc}%
I_m & 0\\
0 & 0_m
\end{array}
\right]
=
 R_{in} [i\omega] .
\end{equation}
That is, in the passive case the steady-state output state $\rho_{field,g}$ is again the vacuum state.

\subsection{Tensors} \label{sec:tensor}

In this subsection several types of tensors and their associated operations are introduced. Because different channels may have different numbers of photons, fibers of the tensors may thus have different lengths, see e.g., (\ref{eq:S_ij}). Nonetheless, with slight abuse of notation, we still call these objects tensors. 

Given positive integers $m$ and $\ell_1,\ldots,\ell_m$, let $\mathbb{C}^{m\times(\ell_1,\ldots,\ell_m)}$ be a space of matrix-like objects, whose element $\mathcal{\xi}$ is of the form
\begin{equation*} \label{eq:xi_0}
\mathcal{\xi}
=
\left[
\begin{array}{ccc}%
\mathcal{\xi}^{11} & \cdots & \mathcal{\xi}^{1\ell_1}\\
\vdots & \ddots & \vdots\\
\mathcal{\xi}^{m1} & \cdots & \mathcal{\xi}^{m\ell_m}
\end{array}
\right]  .
\end{equation*}
In this paper $\xi$ is used to represent $m$-channel multi-photon input states with $\ell_j$ denoting the photon number in the $j$-th channel, $j = 1,\ldots,m$. Because channels may have different numbers of photons,  $\ell_1,\ldots,\ell_m$ may not equal each other. Nonetheless in the paper we still call $\mathcal{\xi}$ a matrix. Next we define a tensor space $\mathbb{C}^{m\times m\times(\ell_1,\ldots,\ell_m)}$, whose elements $\mathscr{S}$ are defined in the following way: For each $i,j=1,\ldots,m$, the model-3 fiber is
\begin{equation} \label{eq:S_ij}
\mathscr{S}_{ij:}
=
\left[
\begin{array}{c}
\mathscr{S}_{ij1}\\
\vdots\\
\mathscr{S}_{ij\ell_{j}}
\end{array}
\right]  \in \mathbb{C}^{\ell_j}.
\end{equation}
That is, when the first two indices $i,j$ are fixed, we have a vector
of dimension $\ell_j$. $\mathscr{S}$ looks
like a 3-way tensor (\cite{KB09}), but its mode-3 fibers may have different
dimensions for different $j$. Nevertheless, in this paper we still call $\mathscr{S}$ a 3-way tensor and $\mathbb{C}^{m\times m\times(\ell_1,\ldots,\ell_m)}$ a space of 3-way tensors over the field of complex numbers.  Given a matrix
$\mathcal{\xi}\in\mathbb{C}^{m\times(\ell_1,\ldots,\ell_m)}$, we may represent it as a 3-way tensor $\mathscr{\xi}^\uparrow \in \mathbb{C}^{m\times m\times(\ell_1,\ldots,\ell_m)}$, by defining horizontal slices to
be
\begin{equation} \label{eq:matrix_to_tensor}
\mathscr{\xi}_{i::}^\uparrow
=
\left[
\begin{array}{l}
\left. \begin{array}{ccc}
  0 & \cdots & 0 \\
  \vdots & \ddots & \vdots \\
  0 & \cdots & 0
\end{array}\right\}_{i-1} \\
\begin{array}{ccc}
\mathscr{\xi}^{i1} & \cdots & \mathscr{\xi}^{i\ell_i}
\end{array} \\
\left. \begin{array}{ccc}
  0 & \cdots & 0 \\
  \vdots & \ddots & \vdots\\
  0 & \cdots & 0
\end{array} \right\}_{m-i}
\end{array}
\right]\in \mathbb{C}^{m\times(\ell_1,\ldots,\ell_m)}, \ \ \forall i=1,\ldots,m.
\end{equation}
This update turns out to be very useful because the output state of a quantum passive linear system driven by an $m$-channel multi-photon state encoded by a matrix $\xi\in\mathbb{C}^{m\times(\ell_1,\ldots,\ell_m)}$ can be characterized by a tensor in $\mathbb{C}^{m\times m\times(\ell_1,\ldots,\ell_m)}$, see Sec. \ref{sec:output_state}.

We adopt notations in \cite{KB09}. For each $j=1,\ldots,m$
and $k=1,\ldots,\ell_j$,
\begin{equation*}
\mathscr{S}_{:jk}
=
\left[
\begin{array}{c}
\mathscr{S}_{1jk}\\
\vdots\\
\mathscr{S}_{mjk}%
\end{array}
\right]  \in\mathbb{C}^m
\end{equation*}
is mode-1 (column) fiber. $\mathscr{S}_{i::}$ and $\mathscr{S}_{:j:}$ are respectively horizontal and lateral slices (matrices) of the form
\begin{equation*}
\mathscr{S}_{i::}
=
\left[
\begin{array}{c}
\mathscr{S}_{i1:}^T\\
\vdots\\
\mathscr{S}_{im:}^T%
\end{array}
\right]  \in\mathbb{C}^{m\times(\ell_1,\ldots,\ell_m)}, ~~
\mathscr{S}_{:j:}%
=\left[
\begin{array}{ccc}%
\mathscr{S}_{:j1} & \cdots & \mathscr{S}_{:j\ell_j}%
\end{array}
\right]  \in\mathbb{C}^{m\times \ell_j},  ~~ \forall i,j=1,\ldots,m.
\end{equation*}
Finally, let $\mathscr{C}\in\mathbb{C}^{m\times(\ell_1,\ldots,\ell_m)\times(\ell_1,\ldots,\ell_m)}$ be a 3-way tensor. We say that $\mathscr{C}$ is partially Hermitian in modes 2 and 3 if all the horizontal slices are Hermitian matrices. That is, for all $i=1,\ldots,m$, the horizontal slices $\mathscr{C}_{i::}\in\mathbb{C}^{\ell_i\times \ell_i}$ satisfy $\mathscr{C}_{i::}^\dag=\mathscr{C}_{i::}$. This is a natural extension of the concept {\it partially symmetric} discussed in (\cite{KB09}) to the complex domain.

In what follows we define operations associated to these tensors.  Given 3-way tensors
$\mathscr{S}(t),\mathscr{T}(r) \in\mathbb{C}^{m\times m\times(\ell_1,\ldots,\ell_m)}$ and partially Hermitian tensor $\mathscr{C}\in\mathbb{C}^{m\times(\ell_1,\ldots,\ell_m)\times(\ell_1,\ldots,\ell_m)}$,
we define a matrix $\mathscr{S}(t)\circledast
\mathscr{T}(r) \in\mathbb{C}^{m\times m}$ whose (i,k)-th entry is%
\begin{equation}
\left(\mathscr{S}(t)\circledast\mathscr{T}(r)\right)_{ik}:=%
\sum\limits_{j=1}^m
\frac{1}{N_{\ell_j}}\sum\limits_{\beta=1}^{\ell_j}
\sum\limits_{\alpha=1}^{\ell_j}
\mathscr{C}_{j\alpha\beta}\mathscr{S}_{ij\alpha}(t)\mathscr{T}_{kj\beta}(r), \ \ \ \forall i, k = 1,\ldots, m,
\label{eq:tensor}
\end{equation}
where $N_{\ell_j}$ ($j=1,\ldots,m$) are positive scalars. (The physical interpretation of $N_{\ell_j}$ will be given in Sec. \ref{sec:factorable}.)
 It can be verified that%
\begin{equation}
\left(\mathscr{S}(t)\circledast\mathscr{T}(r)\right)^\dag
=
\mathscr{T}(r)^\#\circledast\mathscr{S}(t)^\#.
\label{tensor_1_b}
\end{equation}
In this paper, we call $\mathscr{C}$ a ``core tensor'' for the operation $\circledast$.
According to (\ref{eq:tensor}) and the definition of
$\mathscr{\xi}^\uparrow$ in (\ref{eq:matrix_to_tensor}), we have
\begin{eqnarray}
  \mathrm{diag}_{j=1,\ldots,m}(\frac{1}{N_{\ell_j}}
\sum\limits_{i,k=1}^{\ell_j}\mathscr{C}_{jik}\mathscr{\xi}^{ji}(r)^\ast\mathscr{\xi}^{jk}(t)) &=& \mathscr{\xi}^\uparrow(r)^\#\circledast\mathscr{\xi}^\uparrow(t),  \label{eq:up_2a}
\end{eqnarray}
Given a matrix function $E(t)\in\mathbb{C}^{m\times m}$ and a $3$-way tensor $\mathscr{S}(t)\in
\mathbb{C}^{m\times m\times(\ell_1,\ldots,\ell_m)}$, define $\mathscr{T}\in
\mathbb{C}^{m\times m\times(\ell_1,\ldots,\ell_m)}$ whose $(i,j,k)$-th element is
\[
\mathscr{T}_{ijk}(t) := \sum_{r=1}^m\int_{-\infty}^\infty E^{ir}(t-\tau)\mathscr{S}_{rjk}(\tau)d\tau.
\]
In compact form we write
\begin{equation*}
\mathscr{T}= \mathscr{S} \times_1 E, \label{eq:tensor2}
\end{equation*}
where $\times_1$ is called \emph{1-mode matrix product} (\cite[Sec. 2.5]{KB09}).

Given two matrices $E(t),F(t)\in\mathbb{C}^{m\times m}$ and two
tensors $\mathscr{S}(t),\mathscr{T}(t)\in\mathbb{C}^{m\times m\times
(l_1,\ldots,\ell_m)}$, define
\begin{equation} \label{times_one}
\Delta(\mathscr{S},\mathscr{T}) \times_1 \Delta(E,F)
:=
\Delta(\mathscr{S}\times_1 E + \mathscr{T}^\# \times_1 F, \mathscr{T}\times_1 E + \mathscr{S}^\# \times_1 F).
\end{equation}
That is, the operation $\times_1$  is performed block-wise. This operation is useful in studying the output state of a quantum linear system driven by a multi-channel multi-photon input state.

Finally, we define another type of operations between matrices and tensors. Let $E(t)=(E^{jk}(t))\in\mathbb{C}^{m\times m}$ be the transfer function of the underlying quantum linear system with $m$ input channels. For each $j=1,\ldots,m$, let $\mathscr{V}_j(t_1,\ldots,t_{\ell_j})$ be an $\ell_j$-way $m$-dimensional tensor function that encodes the pulse information of the $j$th input field containing $\ell_j$ photons. Denote the entries of $\mathscr{V}_j(t_1,\ldots,t_{\ell_j})$ by $\mathscr{V}_{j,k_1,\ldots,k_{\ell_j}}(t_1,\ldots,t_{\ell_j})$.  Define an  $\ell_j$-way $m$-dimensional tensor $\mathscr{W}_j$ with entries given by the following multiple convolution
\begin{align*}\label{eq:W}
   & \mathscr{W}_{j,r_1,\ldots,r_{\ell_j}}(t_1,\ldots,t_{\ell_j}) \\
  =\sum_{k_1,\ldots,k_{\ell_j}=1}^m & \int_{-\infty}^\infty \cdots \int_{-\infty}^\infty E^{r_1k_1}(t_1-\iota_1)\cdots E^{r_{\ell_j}k_{\ell_j}}(t_{\ell_j}-\iota_{\ell_j})\mathscr{V}_{j,k_1,\ldots,k_{\ell_j}}(\iota_1,\ldots,\iota_{\ell_j})
 d\iota_1\ldots d_{\ell_j} \nonumber
\end{align*}
for all $1\leq r_1,\ldots,r_{\ell_j}\leq m$. 
In compact form we write
\begin{equation}\label{eq:VW_j}
\mathscr{W}_j = \mathscr{V}_j   \times_1 E  \times_2 \cdots  \times_{\ell_j} E , \ \ \ \forall j=1,\ldots,m,
\end{equation}
cf. \cite[Sec. 2.5]{KB09}. More discussions on tensors will be given in Section \ref{sec:active}.

\section{The factorizable case}\label{sec:factorable}
In this section we study how a quantum linear system responds to a factorizable multi-photon input state, here the word ``factorizable'' means that photons in each input channel are not statistically correlated. The single-channel and multi-channel multi-photon input states are defined in Subsections \ref{subsec:single-channel} and \ref{sec:multi_photon_def} respectively, output covariance functions and intensities are presented in Subsection \ref{subsec:covariance_intensity}, while the output states are derived in Subsection \ref{sec:output_state}.

\subsection{Single-channel multi-photon states}\label{subsec:single-channel}
In this subsection  single-channel $\ell$-photon states are defined and their statistical properties are discussed.

For any given positive integer $\ell$ and real numbers $t_1,\ldots,t_\ell$, let $P(t_1,\ldots,t_\ell)$ be a permutation of the numbers $t_1,\ldots,t_\ell$. Denote the set of all such
permutations by $S_\ell$. For arbitrary functions $\xi_1(t),\ldots,\xi_\ell(t)$ defined on the real line,
define
\begin{equation}
N_\ell :=
\sum_{P\in S_\ell}\int_\ell
\xi_\ell(t_\ell)^\ast\cdots\xi_1(t_1)^\ast\xi_1(P(t_1))\cdots\xi_\ell(P(t_\ell))dt_{1\to\ell},
\label{normalization}
\end{equation}
provided the above multiple integral converges (this is always assumed in the
paper). The subscript ``$\ell$'' in $N_\ell$ indicates the
number of photons. It can be shown that $N_\ell>0$. A single-channel
\emph{continuous-mode} $\ell$-photon state $|\psi_\ell\rangle $ is defined via
\begin{equation}
|\psi_\ell\rangle
:=
\frac{1}{\sqrt{N_\ell}}\prod\limits_{k=1}^\ell B^\ast(\xi_k)|0\rangle ,
\label{state}
\end{equation}
where $B^\ast(\xi_k) :=\int_{-\infty}^\infty b^\ast(t)\xi_k(t)dt $, ($k=1,\ldots,m$.) Because $|\psi_\ell\rangle$ is a product of single integrals, there is no correlation among the photons. This type of multi-photon states is therefore called \emph{factorizable} photon states. It can be shown that%
\begin{equation*} \label{inner-product}
\langle0| \prod\limits_{i=1}^\ell B(\xi_i)\prod\limits_{k=1}^\ell B^\ast(\xi_k)|0\rangle
=
\sum_{P\in S_\ell}\int_\ell
\xi_\ell(t_\ell)^\ast\cdots\xi_1(t_1)^\ast \xi_1(P(t_1))\cdots\xi_\ell(P(t_\ell))
dt_{1\to\ell}
=
N_\ell.
\end{equation*}
Thus $\langle\psi_\ell|\psi_\ell\rangle=1$. That is, $|\psi_\ell\rangle $ is normalized.

When $\ell=1$, $N_1=\int_{-\infty}^\infty |\xi_1(t)|^2dt$, $|\psi_1\rangle $ is a single-photon state,
(\cite[(6.3.4)]{RL00}; \cite[(9)]{Milburn08}). On the other hand, when
$\xi_1(t)\equiv\cdots\equiv\xi_\ell(t)\equiv\xi(t)$ and
$\int_{-\infty}^\infty |\xi(t)|^2dt=1$, the input light field contains $\ell$ indistinguishable photons; such
states are called \textit{continuous-mode} Fock states which have been intensely studied, in e.g., \cite[(3)]{GEP+98}; \cite[(13)]{BCB+12}.

For convenience, define a matrix $\mathcal{C}\in\mathbb{C}^{\ell\times \ell}$ whose entries are
\begin{equation}
\mathcal{C}^{ik}
=
\left\langle 0\right\vert
\prod\limits_{\alpha=1,\alpha\neq i}^\ell B(\xi_{\alpha})\prod\limits_{\beta=1,\beta\neq k}^\ell B^\ast(\xi_{\beta})\left\vert 0\right\rangle ,\ \ell\geq2. \label{C2}
\end{equation}
Clearly, $\mathcal{C} = \mathcal{C}^\dag$.

\begin{lem}\label{lem:Nl}
$\mathcal{C}^{ik}$ defined in (\ref{C2}) satisfies
\[
\sum_{k=1}^\ell \mathcal{C}^{ik}\int_{-\infty}^\infty \xi_{i}(t)^\ast\xi_k(t)dt = N_\ell, \ \ \ \forall i=1,\ldots,\ell.
\]
\end{lem}

In what follows we study statistical properties of the $\ell$-photon state $|\psi_\ell\rangle$.  It is easy to show that for all $t\geq t_0$, $\langle\psi_\ell |b(t)|\psi_\ell \rangle=0$. That is, the field has zero average field amplitude. The following result summarizes the second-order statistical information of the $\ell$-photon state $|\psi_\ell\rangle$.

\begin{lem}\label{lem:R_n}
Let $\bar{n}(t)$ denote the field intensity with respect to the state $|\psi_\ell\rangle $, namely,
\begin{equation*}
\bar{n}(t)= \langle\psi_\ell|b^\ast(t)b(t)|\psi_\ell\rangle .
\end{equation*}
(In quantum optics, the second-order moment $\bar{n}(t)$ is the count rate (\cite{GZ00}).)  Moreover, let the field covariance function be
\begin{equation*}
R(t,r)
=
\langle\psi_\ell|\breve{b}(t)\breve{b}^\dag(r)|\psi_\ell\rangle,
\label{cor_0}
\end{equation*}
as given by (\ref{eq:R}). Then we have
\begin{eqnarray}
R(t,r)&=&\delta(t-r)\left[
\begin{array}{cc}%
1 & 0\\
0 & 0
\end{array}
\right]  +\frac{1}{N_\ell} \sum_{i=1}^\ell  \sum_{k=1}^\ell \Delta\left(\mathcal{C}^{ik}\xi_k(t)\xi_i(r)^\ast,0\right) , 
\label{cor-3}\\
\bar{n}(t)
&=&
\frac{1}{N_\ell} \sum_{i=1}^\ell \sum_{k=1}^\ell \mathcal{C}^{ik}\xi_i(t)^\ast\xi_k (t). \label{bar_n_in}%
\end{eqnarray}
\end{lem}

\textbf{Proof.}~
Clearly,
\begin{equation}
R(t,r)
=
\delta(t-r)\left[
\begin{array}{cc}%
1 & 0\\
0 & 0
\end{array}
\right]
+
\Delta(\langle\psi_\ell|b^\ast(r)b(t)|\psi_\ell\rangle, \langle\psi_\ell|b(t)b(r)|\psi_\ell\rangle). \label{cor}
\end{equation}
 Observing that%
\begin{equation}
b(t)|\psi_\ell\rangle
=
\frac{1}{\sqrt{N_\ell} }\sum_{k=1}^\ell \xi_k(t)
\prod\limits_{r=1,r\neq k}^\ell B^\ast(\xi_{r})|0\rangle ,
\label{b_in}
\end{equation}
we have
\begin{equation}
\langle \psi_\ell |b^\ast(r)b(t)|\psi_\ell\rangle
=
\frac{1}{N_\ell} \sum_{i=1}^\ell \sum_{k=1}^\ell \mathcal{C}^{ik}\xi_i(r)^\ast\xi_{k}(t), \ \  \langle \psi_\ell|b(t)b(r)|\psi_\ell\rangle  =0.
\label{b*b}
\end{equation}
Substituting (\ref{b*b}) into (\ref{cor}) establishes (\ref{cor-3}). Finally, because  $\bar{n}(t)$ is the 2-by-2 entry of $R(t,t)$, (\ref{bar_n_in}) follows (\ref{cor-3}).

In particular, for the single-photon case, the field covariance function is
\begin{equation}
R(t,r)=\delta(t-r)\left[
\begin{array}{cc}%
1 & 0\\
0 & 0
\end{array}
\right]  +\Delta\left(\xi_1(t)\xi_1(r)^\ast,0\right)  ,
\label{cor-4}%
\end{equation}
which is the same as \cite[(35)]{ZJ12}. 

\emph{Remark 4.} According to Lemma \ref{lem:R_n}, the $\ell$-photon state $|\psi_\ell\rangle$ is not Gaussian; it may not be stationary either. So, its first and second order moments cannot provide all statistical information of the input field.

\subsection{Multi-channel multi-photon states} \label{sec:multi_photon_def}

In this subsection multi-channel multi-photon states are defined.

Let there be $m$ input field channels. For the j-th field channel, let $\ell_j$ be
the number of photons ($j=1,\ldots,m$). Similar to (\ref{state}), define
the $j$-th channel $\ell_j$-photon state by%
\begin{equation}
|\Psi_j\rangle
:=
\frac{1}{\sqrt{N_{\ell_j}}}\prod\limits_{k=1}^{\ell_j}
B_j^\ast(\xi^{jk})|0\rangle , \label{Psi_in_j}
\end{equation}
where the subscript ``$j$'' indicates the $j$-th channel, and $\ell_j$ indicates that there are $\ell_j$
photons in this channel. In analog to (\ref{normalization}), for each $j=1,\ldots,m$, the normalization coefficient $N_{\ell_j}$ is defined to be
\begin{equation*}
N_{\ell_j}
:=
\sum_{P\in S_{\ell_j}}\int_{\ell_j}
\xi^{j\ell_j}(t_{\ell_j})^\ast\cdots\xi^{j1}(t_1)^\ast\xi^{j1}(P(t_1))\cdots\xi^{j\ell_j}(P(t_{\ell_j}))
dt_{1\to\ell_j}.
\label{normalization_2}
\end{equation*}
We define an $m$-channel multi-photon state as
\begin{equation}
|\Psi\rangle
:=
|\Psi_1\rangle \otimes |\Psi_2\rangle \otimes\cdots\otimes |\Psi_m\rangle
=
\prod\limits_{j=1}^m
\frac{1}{\sqrt{N_{\ell_j}}}\prod\limits_{k=1}^{\ell_j}
B_j^\ast(\xi^{jk}) |0^{\otimes m}\rangle . 
\label{Psi_in}
\end{equation}
In particular, for each $j=1,\ldots,m$, if $\xi^{j1}(t)\equiv\cdots\equiv\xi^{j\ell_j}(t)$,
then (\ref{Psi_in}) defines a multi-channel continuous-mode Fock state,
see eg., \cite[(D1)]{BCB+12}.

\subsection{Output covariance functions and intensities}\label{subsec:covariance_intensity}
In this subsection analytical forms of output covariance functions $R_{\rm out}(t,r)$ and intensities $\bar{n}_{\rm out}(t)$ are presented.

\subsubsection{Steady-state output covariance function}
In this part we derive an explicit expression of $R_{\rm out}(t,r)$ when the
input is in the multi-channel multi-photon state $|\Psi\rangle $ defined in (\ref{Psi_in}). 

We first introduce some notation. Define a 3-way tensor $\mathscr{C}\in\mathbb{C}^{m\times(\ell_1,\ldots,\ell_m)\times(\ell_1,\ldots,\ell_m)}$, whose elements are
\begin{equation}
\mathscr{C}_{jik}:
=
\left\langle0\right\vert
\prod\limits_{\alpha=1,\alpha\neq i}^{\ell_j}
B_j(\xi^{j\alpha})\prod\limits_{\beta=1,\beta\neq k}^{\ell_j}
B_j^\ast(\xi^{j\beta})\left\vert 0\right\rangle ,
\ \ \forall j=1,\ldots,m, \mathrm{and}
\  i,k = 1,\ldots, \ell_j.
\label{C2c}
\end{equation}
Clearly, $\mathscr{C}$ is partially Hermitian, that is, $\mathscr{C}_{j::}=\mathscr{C}_{j::}^\dag \in \mathbb{C}^{\ell_j\times\ell_j}$, ($j=1, \ldots, m$). 
Similar to (\ref{b*b}), for each $j=1,\ldots,m$,
\[
\left\langle\Psi|b_j(t)b_j^\ast(r)|\Psi\right\rangle
=
\delta(t-r)+\frac{1}{N_{\ell_j}}\sum\limits_{i=1}^{\ell_j}
\sum\limits_{k=1}^{\ell_j} \mathscr{C}_{jik}\xi^{ji}(r)^\ast\xi^{jk}(t) .
\]
Consequently, the
input covariance function is
\begin{eqnarray}
 R(t,r) &  = &
\delta(t-r)\left[
\begin{array}{cc}%
I_m & \\
& 0_m%
\end{array}
\right]
+\left[
\begin{array}{cc}%
\xi^\uparrow(r)^\#\circledast\xi^\uparrow(t) & \\
& \left(\xi^\uparrow(r)^\#\circledast\xi^\uparrow(t)\right)
^{\dag}%
\end{array}
\right]. \label{eq:R_in}
\end{eqnarray}
For state $|\Psi\rangle$ defined in (\ref{Psi_in}), let $\xi^\uparrow$ be the 3-way tensor defined via (\ref{eq:matrix_to_tensor}). Then we can  define 3-way tensors $\mathbf{\eta}^-,\mathbf{\eta}^+\in\mathbb{C}^{m\times m \times(\ell_1,\ldots,\ell_m)}$ by
\begin{equation}
\Delta\left(  \mathbf{\eta}^-,\mathbf{\eta}^+\right)
:=
\Delta\left(\xi^\uparrow,\mathbf{0}\right) \times_1  g_G,   \label{eta_0}
\end{equation}
where the operation  $\times_{1}$ has been defined in (\ref{times_one}).
 For example, for a single-channel $\ell$-photon input state defined in (\ref{state}), equation (\ref{eta_0}) yields
\begin{equation}
\eta_k^-(t)
:=
\int_{-\infty}^\infty g_{G^-}(t-\tau)\xi_k(\tau)d\tau, \ \
\eta_k^+(t)
:=
\int_{-\infty}^\infty g_{G^+}(t-\tau)\xi_k(\tau)^\ast d\tau ,  \ k=1,2,\ldots \ell.
\label{eq:single_channel}%
\end{equation}

\begin{thm}\label{thm:covariance-transfer_photon}
Assume that the quantum linear system $G$ is asymptotically stable. Let the input field have covariance $R(t,r)$ be that 
given in (\ref{eq:R_in}). Then the steady-state output covariance function is
\begin{align}
R_{\rm out}(t,r)
&=\int_{-\infty}^\infty g_G(t-\tau)\left[
\begin{array}{cc}%
I_m & \\
& 0_m%
\end{array}
\right] g_G(r-\tau)^\dag d\tau 
\label{R_out}\\
& ~~~ +\Delta\left(\left(\eta^-(r)^\#\circledast\eta^-(t)\right),\left(\eta^+(r)\circledast\eta^-(t)\right)\right)^T +
\Delta\left(\eta^+(t)\circledast\eta^+(r)^\#, \eta^+(t)\circledast\eta^-(r)\right)  , \nonumber
\end{align}
where the tensors $\eta^{-}(t)$ and $\eta^{+}(t)$ are given by (\ref{eta_0}), and the core tensor for the operation $\circledast$ is the tensor $\mathscr{C}$ given in (\ref{C2c}).
\end{thm}

\textbf{Proof.}~
(\ref{R_out}) can be derived by substituting (\ref{eq:R_in}) into (\ref{Eq:out_in}) and with the aid of (\ref{tensor_1_b})-(\ref{eq:up_2a}) and (\ref{eta_0}).

\subsubsection{Steady-state output intensity}

For the multi-channel multi-photon input state $|\Psi\rangle$ defined in (\ref{Psi_in}),  the steady-state ($t_0\to-\infty$) intensity of the output field is
\begin{equation}
\bar{n}_{\rm out}(t)
=
\left\langle \phi\Psi|b_{\rm out}^\#(t)b_{\rm out}^T(t)|\phi\Psi\right\rangle .  \label{n_out2}
\end{equation}
Because $\bar{n}_{\rm out}(t)$ is the 2-by-2 block of $R_{\rm out}(t,t)$, the following result is an immediate consequence of Theorem \ref{thm:covariance-transfer_photon}.

\begin{cor}\label{thm:n-out-2}
Assume the quantum linear system $G$ is asymptotically stable. The steady-state ($t_0\to-\infty$) intensity $\bar{n}_{\rm out}(t)$ of the output field defined in (\ref{n_out2}), of the system $G$ driven by the $m$-channel multi-photon input field $|\Psi\rangle$ defined in (\ref{Psi_in}), is given by%
\begin{equation}
\bar{n}_{\rm out}(t)
=
\int_{-\infty}^\infty g_{G^+}(t)^\# g_{G^+}(t)^T dt
+\left(\eta^+(t)\circledast\eta^+(t)^\#\right)^T
+\eta^-(t)^\#\circledast\eta^-(t),
\label{eq:nout8}
\end{equation}
where $\eta^{-}(t)$ and $\eta^{+}(t)$ are given by (\ref{eta_0}), and the core tensor for the operation $\circledast$ is given in (\ref{C2c}).
\end{cor}

\subsection{Steady-state output state} \label{sec:output_state}

The preceding subsections studied the first  and second order moments of output fields of quantum linear systems driven by multi-photon states. Because the output states are in general not Gaussian, these moments cannot provide the complete information of output fields. In this subsection we derive the analytic form of output states.

A multi-channel continuous-mode multi-photon state $|\Psi\rangle $ defined in (\ref{Psi_in}) is parameterized by the functions $\xi^{jk}(t)$, each of which has two indices $j$ and $k$. The index $j$ (from $1$ to $m$) indicates the
$j$-th input channel, while the index $k$ (from $1$ to $\ell_j$ for each given $j$) is used to count
the number of photons in each channel. On the other hand, according to (\ref{eta_0}), the steady-state output covariance function (in (\ref{R_out})) and intensity $\bar{n}_{\rm out}(t)$ (in (\ref{eq:nout8})) of the linear quantum system $G$ driven by $|\Psi\rangle$ are parameterized by tensor functions $\eta_{ijk}^-(t)$ and
$\eta_{ijk}^+(t)$, each of which has three indices $i,j,k$. Formally, the
index $i$ indicates that each output channel is a linear combination of the
input channels. Interestingly, let $\xi^- = \xi^\uparrow$ (c.f. (\ref{eq:matrix_to_tensor})), that is,
\[
\xi_{ijk}^-(t)
:=
\left\{
\begin{array}{cl}%
0, & i\neq j,\\
\xi^{jk}(t), & i=j.
\end{array}
\right.
\]
Define further $\xi^+ \in\mathscr{C}^{m\times m\times(\ell_1,\ldots,\ell_m)}$ to be a zero tensor. Then the $m$-channel multi-photon state $|\Psi\rangle$ can be re-written as%
\begin{equation}
|\Psi\rangle
=
\prod\limits_{j=1}^m \frac{1}{\sqrt{N_{\ell_j}}}\prod\limits_{k=1}^{\ell_j}
\sum_{i=1}^m (B_i^\ast(\xi_{ijk}^-)-B_i(\xi_{ijk}^+))|0^{\otimes m}\rangle .
\label{Psi_in_2}
\end{equation}
Accordingly, (\ref{eta_0}) can be re-written as%
\begin{equation}
\Delta(\eta^-,\eta^+)  = \Delta(\xi^-,\xi^+) \times_1  g_G . \label{eta}
\end{equation}

In light of the above discussion, we derive the steady-state
output state of the quantum linear system $G$ driven by an input state of the
form
\begin{equation*}
\rho_{\xi,R}
=
\prod\limits_{j=1}^m \frac{1}{\sqrt{N_{\ell_j}}}\prod\limits_{k=1}^{\ell_j}\sum_{i=1}^m
(B_i^\ast(\xi_{ijk}^-)-B_i(\xi_{ijk}^+))
\rho_{R}\left(\prod\limits_{j=1}^m \frac{1}{\sqrt{N_{\ell_j}}}\prod\limits_{k=1}^{\ell_j}\sum_{i=1}^m
(B_i^\ast(\xi_{ijk}^-)-B_i(\xi_{ijk}^+))
\right)^\ast,
\label{Psi_in_3}
\end{equation*}
where
\begin{equation}
\xi(t) = \Delta(\xi^-(t),\xi^+(t))  , \label{state:xi}
\end{equation}
with $\xi^-,\xi^+ \in \mathbb{C}^{m\times m\times(\ell_1,\ldots,\ell_m)}$ and $\rho_R$ is a stationary zero-mean Gaussian state with covariance function $R[i\omega]$.

In order for $\rho_{\xi,R}$ to be a valid state, $\xi^-,\xi^+$ and $\rho_R$ have to satisfy certain conditions. We first introduce some notation. Given a tensor $\varphi\in\mathbb{C}^{2m\times m\times(\ell_1,\ldots,\ell_m)}$, define
tensor products consisting of $\ell_j$ vectors, each of dimension $2m$:%
\begin{align*}
M_{\varphi_{:j}}(t_{1\rightarrow \ell_j})
&  :=
\varphi_{:j1}(t_1)\otimes_c\cdots\otimes_c\varphi_{:j\ell_j}(t_{\ell_j}), \ \ \ j=1,\ldots,m,
\\
M_{\varphi_{:j}}^+(t_{1\rightarrow \ell_j})
&  :=
\varphi_{:j\ell_j}(t_1)\otimes_c\cdots\otimes_c\varphi_{:j1}(t_j), \ \ \ j=1,\ldots,m,
\end{align*}
where $\otimes_c$ is the Kronecker product as introduced in the {\it Notations} part. Then define tensor products of the form
\begin{eqnarray*}
 M_\varphi(t_{1\rightarrow \ell_1+\cdots+\ell_m})
:=
\frac{1}{\sqrt{N_{\ell_1}}}M_{\varphi_{:1}}(t_{1\rightarrow \ell_1})\otimes_c%
\cdots\otimes_c\frac{1}{\sqrt{N_{\ell_m}}}M_{\varphi_{:m}}(t_{\ell_1+\cdots+\ell_{m-1}+1\rightarrow \ell_1+\cdots+\ell_m}),
\label{eq:M_xi_3}\\
M_\varphi^+(t_{1\rightarrow \ell_1+\cdots+\ell_m})
:=
 \frac{1}{\sqrt{N_{\ell_m}}}M_{\varphi_{:m}}^+(t_{1\rightarrow \ell_m})\otimes_c%
\cdots\otimes_c\frac{1}{\sqrt{N_{\ell_1}}}M_{\varphi_{:1}}^+(t_{\ell_2+\cdots+\ell_m+1\rightarrow \ell_1+\cdots+\ell_m}). \label{eq:M_xi_4}
\end{eqnarray*}
 Similarly, for the operators $\breve{b}(t)$, define
\begin{equation*}
M_{\breve{b}}(t_{1\rightarrow k}):=\breve{b}(t_1)\otimes_{c}\cdots\otimes_c\breve{b}(t_k), \ \  \forall k\geq 1. \label{eq:M_b}
\end{equation*}
Finally for a matrix $A$, let $A^{\otimes_c^k}:=A\otimes_c\cdots\otimes_c A$
be an $k$-way Kronecker tensor product.

The following equation will be used in Definition \ref{def:F_1}.%
\begin{align}
&  \int_{2\sum_{j=1}^m\ell_j}
(M_\xi^+(t_{1\rightarrow \ell_1+\cdots+\ell_m})^\# \otimes_c
M_\xi(t_{\ell_1+\cdots+\ell_m+1\rightarrow2(\ell_1+\cdots+\ell_m)}))^T J^{\otimes_c^{\ell_1+\cdots+\ell_m}}\nonumber\\
&  \ \ \ \ \ \otimes_c\Theta^{\otimes_c^{\ell_1+\cdots+\ell_m}}
\mathrm{Tr}[\rho_R M_{\breve{b}}(t_{1\rightarrow2(\ell_1+\cdots+\ell_m)})]
dt_{1\rightarrow2(\ell_1+\cdots+\ell_m)}=1,
\label{eq:innerproduct}
\end{align}
where $\xi$ is given in (\ref{state:xi}) and $\Theta =
[
\begin{array}{cccc}
0 & I; & -I & 0%
\end{array}%
]$ as introduced in the \emph{Notations} part.

\begin{defn} \label{def:F_1} 
A state $\rho_{\xi,R}$ is said to be a
\emph{photon-Gaussian} state if it belongs to the set
\begin{align}
\mathcal{F}_0  & :=\left\{
\rho_{\xi,R}
=
\prod\limits_{j=1}^m \frac{1}{\sqrt{N_{\ell_j}}}\prod\limits_{k=1}^{\ell_j}\sum_{i=1}^m
(B_i^\ast(\xi_{ijk}^-)-B_i(\xi_{ijk}^+))
\rho_{R}
\right.  
\label{class_F_0}\\
& \left. ~ \times\left(\prod\limits_{j=1}^m \frac{1}{\sqrt{N_{\ell_j}}}\prod\limits_{k=1}^{\ell_j}\sum_{i=1}^m
(B_i^\ast(\xi_{ijk}^-)-B_i(\xi_{ijk}^+))
\right)^\ast :\xi\mathrm{~and\mathrm{~}\rho_R~satisfy~} (\ref{eq:innerproduct})\right\} . 
\nonumber
\end{align}
\end{defn}

\emph{Remark 5}. Clearly, the $m$-channel multi-photon state $|\Psi\rangle$ defined in (\ref{Psi_in_2}) belongs to $\mathcal{F}_0$.

\begin{prop}\label{prop:normalization}
The photon-Gaussian states $\rho_{\xi,R}\in\mathcal{F}_0$ are normalized, that is $\mathrm{Tr}[\rho_{\xi,R}]=1$.
\end{prop}

\textbf{Proof.}~
For each $j=1,\ldots,m$, define $\xi_{:j}(t)\in\mathbb{C}^{2m\times2\ell_j}$ by
\begin{equation*}
\xi_{:j}(t):=\Delta\left(  \xi_{:j:}^-(t),\xi_{:j:}^+(t)\right)
=
[\xi_{:j1}(t)\ \cdots\ \xi_{:j(2\ell_j)}(t)], \label{eq:xi}
\end{equation*}
where
\[
\xi_{:jk}(t)
=
\left[
\begin{array}{c}%
\xi_{:jk}^-(t)\\
\xi_{:jk}^+(t)^\#
\end{array}
\right] \in \mathbb{C}^{2m}  , \ \ \ \forall k=1,\ldots,2\ell_j.
\]
It can be shown that
\begin{eqnarray*}
&&\prod\limits_{j=1}^m
\frac{1}{\sqrt{N_{\ell_j}}}\prod\limits_{k=1}^{\ell_j}
\sum_{i=1}^m\left(B_i^\ast(\xi_{ijk}^-)-B_i(\xi_{ijk}^+)\right)
\\
&=&
\prod\limits_{j=1}^m
\frac{1}{\sqrt{N_{\ell_j}}}%
\prod\limits_{k=1}^{\ell_j}
\int_{-\infty}^\infty
\left[
\begin{array}{cc}
-\xi_{:jk}^+(t)^\dag & \xi_{:jk}^-(t)^T%
\end{array}
\right]  \breve{b}(t)dt
 \\
&=&%
\prod\limits_{j=1}^m
\frac{1}{\sqrt{N_{\ell_j}}}%
\prod\limits_{k=1}^{\ell_j}
\int_{-\infty}^\infty\xi_{:jk}(t)^T(\Theta\breve{b}(t))dt
\\
&=&  %
\prod\limits_{j=1}^m
\frac{1}{\sqrt{N_{\ell_j}}}\int_{\ell_j}\left(\xi_{:j1}(t_1)\otimes_c\xi^{:j2}(t_2)
\otimes_c\cdots\otimes_c\xi_{:j\ell_j}(t_{\ell_j})\right)^T (\Theta\breve{b}(t_1))\otimes(\Theta\breve{b}(t_2))\cdots\otimes(\Theta\breve{b}(t_{\ell_j}))
dt_{1\rightarrow \ell_j}
\\
&=&
\int_{\ell_1+\cdots+\ell_m}\left(M_{\xi_{:1}}(t_{1\rightarrow \ell_1})\otimes_c\cdots\otimes_c M_{\xi_{:m}}(t_{\ell_1+\cdots+\ell_{m-1}+1\rightarrow \ell_1+\cdots+\ell_m})\right)^T
(\Theta\breve{b}(t_1))\otimes\cdots\otimes(\Theta\breve{b}(t_{\ell_1+\cdots+\ell_m}))dt_{1\rightarrow \ell_1+\cdots+\ell_m}
\\
&=&
\int_{\ell_1+\cdots+\ell_m} M_{\xi}(t_{1\rightarrow \ell_1+\cdots+\ell_m})^T\Theta^{\otimes_c^{\ell_1+\cdots+\ell_m}}M_{\breve{b}}(t_{1\rightarrow \ell_1+\cdots+\ell_m})
dt_{1\rightarrow \ell_1+\cdots+\ell_m},
\end{eqnarray*}
 where $\Theta =
[
\begin{array}{cccc}
0 & I; & -I & 0%
\end{array}%
]$ as introduced in the \emph{Notations} part.
Thus, that $\mathrm{Tr}[\rho_{\xi,R}]=1$ is equivalent to that (\ref{eq:innerproduct}) holds. The proof is completed.

The following result is the main result of this subsection.

\begin{thm}\label{thm:output_state}
Suppose that the linear quantum system $G$ is asymptotically stable. Then the steady-state output state of $G$ driven by a state $\rho_{\xi,R}\in\mathcal{F}_0$ is
\begin{equation}
\rho_{\eta,R_{\rm out}}=
\prod\limits_{j=1}^m
\frac{1}{\sqrt{N_{\ell_j}}}\prod\limits_{k=1}^{\ell_j}
\sum_{i=1}^m(B_i^\ast(\eta_{ijk}^-)-B_i(\eta_{ijk}^+))
\rho_{\rm field, g}
\left(\prod\limits_{j=1}^{m}
\frac{1}{\sqrt{N_{\ell_j}}}\prod\limits_{k=1}^{\ell_j}
\sum_{i=1}^m(B_i^\ast(\eta_{ijk}^-)-B_i(\eta_{ijk}^+))\right)^\ast,
\label{Psi_out_4}
\end{equation}
where the 3-way tensors $\eta^-$ and $\eta^+$ are given by (\ref{eta}), and $\rho_{\rm field, g}$ is a stationary zero-mean Gaussian field whose covariance function is
\begin{equation*}
R_{\rm out}[i\omega]=G[i\omega] R[i\omega]G[i\omega]^\dag.
\label{Phi}
\end{equation*}
given by the Gaussian transfer (\ref{eq:R-transfer}) in Theorem \ref{thm:spectral-transfer}. Clearly, $\rho_{\eta,R_{\rm out}}\in\mathcal{F}_0$.
\end{thm}

\textbf{Proof.}~ Let $\rho(t,t_0)$ be the density operator of the composite system. Then
\begin{equation*}
\rho(t,t_0) = U(t,t_0)|\phi\rangle\langle\phi|\otimes\rho_{\xi,R}U(t,t_0)^\ast.
\end{equation*}
We study the steady-state behavior of the state, that is, we assume that the interaction starts in the distant past ($t_0\to-\infty$), and also let $t\to\infty$.
\begin{eqnarray}
&& \rho_\infty
\label{eq:temp4} \\
&:=&
\lim_{\begin{subarray}{c}t_0\rightarrow-\infty \\ t\rightarrow\infty\end{subarray}} \rho(t,t_0)
\nonumber\\
&=&
\lim_{\begin{subarray}{c}t_0\rightarrow-\infty \\ t\rightarrow\infty\end{subarray}}
U(t,t_0)|\phi\rangle\langle\phi|\otimes\rho_{\xi,R}U(t,t_0)^\ast
\nonumber\\
&=&
\lim_{\begin{subarray}{c}t_0\rightarrow-\infty \\ t\rightarrow\infty\end{subarray}}
\prod\limits_{j=1}^m\frac{1}{\sqrt{N_{\ell_j}}}%
\prod\limits_{k=1}^{\ell_j}\sum_{i=1}^m U(t,t_0) \left(I_{sys}\otimes (B_i^\ast(\xi_{ijk}^-)-B_i(\xi_{ijk}^+))\right)
U(t,t_0)^\ast\rho_{\infty,{\rm g}}
\nonumber\\
&&
~~~~ \times\lim_{\begin{subarray}{c}t_0\rightarrow-\infty \\ t\rightarrow\infty\end{subarray}}\left(
\prod\limits_{j=1}^m\frac{1}{\sqrt{N_{\ell_j}}}%
\prod\limits_{k=1}^{\ell_j}\sum_{i=1}^m U(t,t_0) \left(I_{sys}\otimes (B_i^\ast(\xi_{ijk}^-)-B_i(\xi_{ijk}^+))\right)U(t,t_0)^\ast\right)^\ast ,
\nonumber
\end{eqnarray}
where $\rho_{\infty,{\rm g}}$ is given in (\ref{eq:rho_inf_g}). According to Lemmas \ref{lem:G_inv} and \ref{lem:G_inv_2}, we have
\begin{equation} \label{eq:b_minus}
\left[
\begin{array}{c}
b_j^-(t,-\infty) \\
b_j^{-^\ast}(t,-\infty)%
\end{array}%
\right]
=
\left[
\begin{array}{c}
\sum_{k=1}^m\int_{-\infty}^\infty g_{G^-}^{kj}(r-t)^\ast b_k(r)-g_{G^+}^{kj}(r-t)b_k^\ast(r)dr \\
\sum_{k=1}^m\int_{-\infty}^\infty -g_{G^+}^{kj}(r-t)^\ast b_k(r)+g_{G^-}^{kj}(r-t)b_k^\ast (r)dr%
\end{array}%
\right].
\end{equation}
As a result,
\begin{align}
& \lim_{\begin{subarray}{c}t_0\rightarrow-\infty \\ t\rightarrow\infty\end{subarray}}%
\prod\limits_{j=1}^m
\frac{1}{\sqrt{N_{\ell_j}}}\prod\limits_{k=1}^{\ell_j}
\sum_{i=1}^{m}U\left(t,t_0\right) \left(I_{sys}\otimes \left(B_i^\ast(\xi_{ijk}^-)-B_i(\xi_{ijk}^+)\right)\right)
U\left(t,t_0\right)^\ast
\nonumber\\
& =%
\prod\limits_{j=1}^m
\frac{1}{\sqrt{N_{\ell_j}}}%
\prod\limits_{k=1}^{\ell_j}
\sum_{i=1}^m\int_{-\infty}^\infty\left(\xi_{ijk}^-(t_k)b_i^{-\ast}(t_k,-\infty)dt_k-\xi_{ijk}^+(t_k)^\ast b_i^-(t_k,-\infty)dt_k\right)
\nonumber\\
& =%
\prod\limits_{j=1}^m
\frac{1}{\sqrt{N_{\ell_j}}}\prod\limits_{k=1}^{\ell_j}
\sum_{i=1}^m\int_{-\infty}^\infty\left(-\int_{-\infty}^\infty%
\sum_{r=1}^m g_{G^+}^{ri}(\iota-t_k)^\ast\xi_{ijk}^-(t_k)
dt_k b_r(\iota)d\iota+\int_{-\infty}^\infty\sum_{r=1}^m g_{G^-}^{ri}(\iota-t_k)\xi_{ijk}^-(t_{k})dt_k b_{r}^\ast(\iota)d\iota\right.
\nonumber\\
& \left. \ \ \  -\int_{-\infty}^\infty\sum_{r=1}^m g_{G^-}^{ri}(\iota-t_k)^\ast\xi_{ijk}^+(t_k)^\ast dt_k b_r (\iota)d\iota
+\int_{-\infty}^\infty\sum_{r=1}^m g_{G^+}^{ri}(\iota-t_k)\xi_{ijk}^+(t_k)^\ast dt_k b_r^\ast(\iota)d\iota\right)
\nonumber\\
& =%
\prod\limits_{j=1}^m\frac{1}{\sqrt{N_{\ell_j}}}%
\prod\limits_{k=1}^{\ell_j}
\sum_{i=1}^m\left(B_i^\ast(\eta_{ijk}^-)-B_i(\eta_{ijk}^+)\right),
\label{eq:temp5}
\end{align}
where 3-way tensors $\eta^-$ and $\eta^+$ are given by (\ref{eta}). Substituting (\ref{eq:temp5}) into (\ref{eq:temp4}) we have
\begin{equation*}
\rho_\infty =%
\prod\limits_{j=1}^m
\frac{1}{\sqrt{N_{\ell_j}}}\prod\limits_{k=1}^{\ell_j}
\sum_{i=1}^m(B_i^\ast(\eta_{ijk}^-)-B_i(\eta_{ijk}^+))
\rho_{\infty,{\rm g}}
\left(\prod\limits_{j=1}^{m}
\frac{1}{\sqrt{N_{\ell_j}}}\prod\limits_{k=1}^{\ell_j}
\sum_{i=1}^m(B_i^\ast(\eta_{ijk}^-)-B_i(\eta_{ijk}^+))\right)^\ast.
\end{equation*}
Tracing out the system part, (\ref{Psi_out_4}) is obtained. The proof is completed.

Restricted to the single-channel case, we have

\begin{cor} \label{thm:output_state_single_channel}
Assume that the quantum linear system $G$ is asymptotically stable. If the single-channel $\ell$-photon input state is $\vert\psi_\ell\rangle$ defined in (\ref{state}), the steady-state output state is
\begin{equation}
\rho_{out}
=
\frac{1}{\sqrt{N_\ell} }\prod\limits_{k=1}^\ell (B^\ast(\eta_k^-)-B(\eta_k^+))
\rho_{field,g}
\left(\frac{1}{\sqrt{N_\ell} }\prod\limits_{k=1}^\ell B^\ast(\eta_k^-)-B(\eta_k^+)\right)^\ast, \label{Psi_out_0}%
\end{equation}
where $\eta_k^-(t)$ and $\eta_k^+(t)$ are given by (\ref{eq:single_channel}) and $\rho_{field,g}$ is given in (\ref{eq:rho_field}).
\end{cor}

Specific to the passive case, the steady-state output state is a multi-photon state, as given by the following result.
\begin{cor} \label{cor:output_state_multiple_channel_passive}
Assume that the quantum linear system $G$ is asymptotically stable and passive. The steady-state output state of $G$ driven by the multi-photon state $|\Psi\rangle$ is a pure state
\begin{equation*}
|\Psi_{\rm out}\rangle
=
\prod\limits_{j=1}^m
\frac{1}{\sqrt{N_{\ell_j}}}\prod\limits_{k=1}^{\ell_j}
\sum_{i=1}^m B_i^\ast(\eta_{ijk}^-)|0^{\otimes m}\rangle . 
\end{equation*}
In particular, in the single-channel case, the steady-state output state of $G$ driven by the single-photon state $|\psi_l\rangle$ is
\begin{equation}
|\Psi_{\rm out}\rangle
=
\frac{1}{\sqrt{N_\ell} }\prod\limits_{k=1}^\ell B^\ast(\eta_k^-)|0\rangle, \label{Psi_out_0_passive}%
\end{equation}
\end{cor}

\emph{Example 1: Beamsplitter}. Consider a beamsplitter with parameters $L=0$, $H=0$, and
\begin{equation*} \label{bm}
S
=
\left[
\begin{array}{cc}%
\sqrt{\eta} & \sqrt{1-\eta}\\
-\sqrt{1-\eta} & \sqrt{\eta}%
\end{array}
\right]  , \ \ \ (0<\eta<1).
\end{equation*}
Let each input channel have two photons. As with (\ref{Psi_in}), define an input state to be
$|\Psi\rangle
=%
\prod\limits_{j=1}^2 \frac{1}{\sqrt{N_{2_j}}}\prod\limits_{k=1}^2
B_j^\ast(\xi^{jk})|0^{\otimes2}\rangle$,
where $N_{2_j}
  =
\int_{-\infty}^\infty|\xi^{j1}(t)\vert^2dt\int_{-\infty}^\infty|\xi^{j2}(t)\vert^2dt
+\left\vert\int_{-\infty}^\infty\xi^{j2}(t)^\ast\xi^{j1}(t)dt\right\vert^2$, ($j=1,2$).
According to (\ref{eta}), we have $\eta^-\in\mathscr{C}^{2\times2\times2}$ with elements
\begin{eqnarray*}
\eta_{111}^-(t)=\sqrt{\eta}\xi^{11}(t), \ \ \
\eta_{112}^-(t)=\sqrt{\eta}\xi^{12}(t), \ \ \
\eta_{121}^-(t)=\sqrt{1-\eta}\xi^{21}(t), \ \ \
\eta_{122}^-(t)=\sqrt{1-\eta}\xi^{22}(t),
\\
\eta_{211}^-(t)=-\sqrt{1-\eta}\xi^{11}(t), \ \ \
\eta_{212}^-(t)=-\sqrt{1-\eta}\xi^{12}(t), \ \ \
\eta_{221}^-(t)=\sqrt{\eta}\xi^{21}(t), \ \ \
\eta_{222}^-(t)=\sqrt{\eta}\xi^{22}(t).
\end{eqnarray*}
According to Corollary \ref{cor:output_state_multiple_channel_passive},
\begin{align}
 |\Psi_{\rm out}\rangle
 =&\frac{\eta(1-\eta)}{\sqrt{N_{2_{1}}N_{2_{2}}}}B_{1}^{\ast}(\xi^{11}%
)B_{1}^{\ast}(\xi^{12})B_{1}^{\ast}(\xi^{21})B_{1}^{\ast}(\xi^{22})\left\vert
0^{\otimes4}\right\rangle 
\nonumber\\
&  +\frac{\eta\sqrt{\eta(1-\eta)}}{\sqrt{N_{2_{1}}N_{2_{2}}}}B_{1}^{\ast}%
(\xi^{11})B_{1}^{\ast}(\xi^{12})\left(  B_{1}^{\ast}(\xi^{22})B_{2}^{\ast}%
(\xi^{21})+B_{1}^{\ast}(\xi^{21})B_{2}^{\ast}(\xi^{22})\right)  \left\vert
0^{\otimes4}\right\rangle 
\nonumber\\
&  -\frac{\sqrt{\eta(1-\eta)}(1-\eta)}{\sqrt{N_{2_{1}}N_{2_{2}}}}B_{1}^{\ast
}(\xi^{21})B_{1}^{\ast}(\xi^{22})\left(  B_{1}^{\ast}(\xi^{12})B_{2}^{\ast
}(\xi^{11})+B_{1}^{\ast}(\xi^{11})B_{2}^{\ast}(\xi^{12})\right)  \left\vert
0^{\otimes4}\right\rangle 
\nonumber\\
&  +\frac{\eta^{2}}{\sqrt{N^{21}N_{2_{2}}}}B_{1}^{\ast}(\xi^{11})B_{1}^{\ast
}(\xi^{12})B_{2}^{\ast}(\xi^{21})B_{2}^{\ast}(\xi^{22})\left\vert 0^{\otimes
4}\right\rangle 
\nonumber \\
& +\frac{(1-\eta)^{2}}{\sqrt{N_{2_{1}}N_{2_{2}}}}B_{1}^{\ast
}(\xi^{21})B_{1}^{\ast}(\xi^{22})B_{2}^{\ast}(\xi^{11})B_{2}^{\ast}(\xi
^{12})\left\vert 0^{\otimes4}\right\rangle 
\nonumber\\
&  -\frac{\eta(1-\eta)}{\sqrt{N_{2_{1}}N_{2_{2}}}}\left(  B_{1}^{\ast}%
(\xi^{11})B_{2}^{\ast}(\xi^{12})+B_{1}^{\ast}(\xi^{12})B_{2}^{\ast}(\xi
^{11})\right)  \left(  B_{1}^{\ast}(\xi^{21})B_{2}^{\ast}(\xi^{22}%
)+B_{1}^{\ast}(\xi^{22})B_{2}^{\ast}(\xi^{21}\right)  \left\vert 0^{\otimes
4}\right\rangle 
\nonumber\\
&  -\frac{\eta\sqrt{\eta(1-\eta)}}{\sqrt{N_{2_{1}}N_{2_{2}}}}\left(
B_{1}^{\ast}(\xi^{11})B_{2}^{\ast}(\xi^{12})+B_{1}^{\ast}(\xi^{12})B_{2}%
^{\ast}(\xi^{11})\right)  B_{2}^{\ast}(\xi^{21})B_{2}^{\ast}(\xi
^{22})\left\vert 0^{\otimes4}\right\rangle 
\nonumber\\
&  +\frac{\sqrt{\eta(1-\eta)}(1-\eta)}{\sqrt{N_{2_{1}}N_{2_{2}}}}\left(
B_{1}^{\ast}(\xi^{21})B_{2}^{\ast}(\xi^{22})+B_{1}^{\ast}(\xi^{22})B_{2}%
^{\ast}(\xi^{21})\right)  B_{2}^{\ast}(\xi^{11})B_{2}^{\ast}(\xi
^{12})\left\vert 0^{\otimes4}\right\rangle 
\nonumber\\
&  +\frac{\eta(1-\eta)}{\sqrt{N_{2_{1}}N_{2_{2}}}}B_{2}^{\ast}(\xi^{11}%
)B_{2}^{\ast}(\xi^{12})B_{2}^{\ast}(\xi^{21})B_{2}^{\ast}(\xi^{22})\left\vert
0^{\otimes4}\right\rangle . 
\label{Psi_out_5}
\end{align}
Assume $\eta=\frac{1}{2}$, that is the system is a balanced beamsplitter. If
$\xi^{11}(t)\equiv\xi^{12}(t)\equiv\xi^{21}(t)\equiv\xi^{22}(t)$
and
$\int_{-\infty}^\infty|\xi^{11}(t)\vert^2dt=1$, then $N_{2_1}=N_{2_2}=2$.
Let $\frac{1}{\sqrt{i!}\sqrt{k!}}|i,k\rangle $ be the state
with $i$ photons in the first channel and $k$ photons in the second channel
respectively, ($i=0,\ldots,4$). (\ref{Psi_out_5}) reduces to
\begin{eqnarray}
|\Psi_{\rm out}\rangle
&  = &
\sqrt{\frac{3}{8}}|4,0\rangle
-\frac{1}{2}|2,0\rangle |0,2\rangle
+\sqrt{\frac{3}{8}}|0,4\rangle .  \label{Psi_out_6}
\end{eqnarray}
(\ref{Psi_out_6}) is the same as (15)  in (\cite{Ou07}). In a similar way, (17) in (\cite{Ou07}) can also be re-produced.

\section{The unfactorizable case}\label{sec:unfactorable}
The factorizable multi-photon states studied in Secion \ref{sec:factorable} form a subclass of more general multi-photon states, e.g., \cite[(58)]{GEP+98} and \cite[Section 2]{BCB+12}. In this section, we study the response of quantum linear systems to general multi-channel multi-photon states where there may exist correlation among photons in channels.

\subsection{More general multi-photon states}\label{subsec:general_photon}
The unfactorizable multi-photon states are defined in this subsection.

If a single channel has $\ell$ photons, a general form of continuous-mode $\ell$-photon state is
\begin{equation}
|\psi_\ell\rangle
=
\frac{1}{\sqrt{N_\ell} }\int_\ell\psi(t_1,\ldots,t_\ell)b^\ast(t_1)\cdots b^\ast(t_\ell)dt_{1\to \ell}|0\rangle , \label{general_Phi}
\end{equation}
where $\psi(t_1,\ldots,t_\ell)$ is a multi-variable function, and
\[
N_\ell=\sum_{P\in S_\ell}\int_\ell \psi(t_1,\ldots,t_\ell) \psi(P(t_1),\ldots,P(t_\ell))dt_{1\to \ell}
\]%
is a normalization parameter, with $P(t_1,\ldots,t_\ell)$ and $S_\ell$ as those defined in subsection \ref{subsec:single-channel}.
In general, for the $m$-channel case, assume the $j$-th channel has $\ell_j$ photons, and the state for this channel is
\begin{equation} \label{eq:general_state_-1}
|\Psi_j\rangle
=
\frac{1}{\sqrt{N_{\ell_j}}}\int_{\ell_j}\Psi_j(t_1,\ldots,t_{\ell_j})b_j^\ast(t_1)\cdots b_j^\ast(t_{\ell_j})
dt_{1 \to \ell_j}|0\rangle .
\end{equation}
Then the state for the $m$-channel input field can be defined as
\begin{equation} \label{eq:general_state_0}
|\Psi\rangle =\prod\limits_{j=1}^m |\Psi_j\rangle = \prod\limits_{j=1}^m \frac{1}{\sqrt{N_{\ell_j}}}\int_{\ell_j}\Psi_j(t_1,\ldots,t_{\ell_j})b_j^\ast(t_1)\cdots b_j^\ast(t_{\ell_j})
dt_{1 \to \ell_j}|0^{\otimes m}\rangle .
\end{equation}

\emph{Remark 6.} In particular, when $\Psi_j(t_1,\ldots,t_{\ell_j}) = \prod_{k=1}^{\ell_j}\Psi^{jk}(t_k)$, ($j=1,\ldots,m$), (\ref{eq:general_state_-1}) reduces to (\ref{Psi_in_j}), and correspondingly (\ref{eq:general_state_0})  reduces to (\ref{Psi_in}), the factorizable case.

\subsection{The passive case} \label{subsec:general_passive}
In this subsection we study the response of the quantum linear passive system $G$ to an  $m$-channel input field in the state $|\Psi\rangle$ defined in (\ref{eq:general_state_0}).

We first rewrite the $m$-channel multi-photon state $|\Psi\rangle$ into an alternative form; this will enable us to present the input and output states in a unified form. For $j=1,\ldots,m$, $i=1,\ldots,\ell_j$, and $k_i=1,\ldots,m$, define
\begin{equation} \label{eq:psi_update}
\Psi_{j,k_1,\ldots,k_{\ell_j}}(\tau_1,\ldots,\tau_{\ell_j})
:= \left\{\begin{array}{cc}
             \Psi_j(\tau_1,\ldots,\tau_{\ell_j}), & k_1=\cdots=k_{\ell_j} = j, \\
             0, & \mathrm{otherwise}.
           \end{array}
   \right.
\end{equation}
Thus  for each $j=1,\ldots,m$ we have an $\ell_j$-way $m$-dimensional tensor, denoted $\Psi_j$. The multi-channel multi-photon state $|\Psi\rangle$ in (\ref{eq:general_state_0}) can be re-written as
\begin{equation*}
|\Psi\rangle
=
\prod_{j=1}^m \frac{1}{\sqrt{N_{\ell_j}}}\sum_{k_1,\ldots,k_{\ell_j}=1}^m\int_{\ell_j} \Psi_{j,k_1,\ldots,k_{\ell_j}}(\tau_1,\ldots,\tau_{\ell_j}) b_{k_1}^\ast(\iota_1)\cdots b_{k_{\ell_j}}^\ast(\iota_{\ell_j})
d\iota_{1\to \ell_j}|0^{\otimes m}\rangle .
\end{equation*}
We define a class of pure states
\begin{align}
\mathcal{F}_1 = \left\{|\Psi\rangle
=
\prod_{j=1}^m \frac{1}{\sqrt{N_{\ell_j}}}\sum_{k_1,\ldots,k_{\ell_j}=1}^m \int_{\ell_j} \Psi_{j,k_1,\ldots,k_{\ell_j}}(\tau_1,\ldots,\tau_{\ell_j})\prod_{i=1}^{\ell_j} b_{k_i}^\ast(\tau_i)
d\tau_{1\to \ell_j}|0^{\otimes m}\rangle : \langle\Psi|\Psi\rangle=1\right\} . \label{eq:F_1}
\end{align}

\begin{thm}\label{thm:passive_multi_channel}
Suppose that the quantum linear system $G$ is asymptotically stable and passive. The steady-state output state of  $G$ driven by a state $|\Psi_{\rm in}\rangle\in \mathcal{F}_1$ is another state $|\Psi_{\rm out}\rangle\in \mathcal{F}_1$ with wave packet transfer
\begin{align*}
\Psi_{{\rm out},j} =  \Psi_{{\rm in},j} \times_1 g_{G^-} \times_2\cdots\times_{\ell_j} g_{G^-} , \ \ \ \forall j=1,\ldots,m,
\label{eq:psi_out}
\end{align*}
where the operation between the matrix and tensor is defined in (\ref{eq:VW_j}).
\end{thm}

Because Theorem \ref{thm:passive_multi_channel} is a special case of Theorem \ref{thm_active_2} for the active case, cf. Remark 9, its proof is omitted.

In particular, for the single-channel case, we have
\begin{cor}\label{thm:general_state}
The steady-state output state of a quantum linear passive system $G$ driven by the $\ell$-photon state $|\psi_\ell\rangle$ in (\ref{general_Phi}) is an $\ell$-photon state
\begin{equation*}
|\psi_{\rm out}\rangle
=
\frac{1}{\sqrt{N_\ell} }\int_\ell\psi_{\rm out}^-(\iota_1,\ldots,\iota_{\ell})b^\ast(t_1)b^\ast(t_2)\cdots b^\ast(t_\ell)
dt_{1\to\ell} |0\rangle ,
\end{equation*}
where the multi-variable function $\psi_{\rm out}^-$ is
\begin{equation*}
\psi_{\rm out}^-(t_1,\ldots,t_\ell) = \int_{-\infty}^\infty \cdots \int_{-\infty}^\infty   g_{G^-}(t_1-\tau_1)\cdots g_{G^-}(t_\ell-\tau_\ell) \psi_\ell (\tau_1,\ldots,\tau_\ell)d\tau_1\cdots d\tau_\ell .
\end{equation*}
\end{cor}

\subsection{The active case}\label{sec:active}

In this subsection we study the response of the quantum linear system $G$ to the  $m$-channel input field in the state $|\Psi\rangle$ defined in (\ref{eq:general_state_0}). Here $G$ is not necessarily passive. In this case, as shown in Sec. \ref{sec:output_state},  $g_{G+}$ contributes to the output states. So the active case is more mathematically involved.

We first introduce some more notations in order to derive the steady-state output state. Define
\begin{equation*} \label{eq:d_2}
\mathrm{sgn}(d_i)
:= \left\{
\begin{array}{ll}
  1, & d_i=1, \\
  0, & d_i=-1,
\end{array}
 \right. \ \ \ \forall i=1,\ldots, \max\{\ell_1,\ldots,\ell_m\} .
\end{equation*}
For each $j=1,\ldots,m$, $i = 1,\ldots, \ell_j$ and $k_i=1,\ldots,m$, define
\begin{equation} \label{eq:psi_update_2}
\Psi_{k_1,\ldots,k_{\ell_j}}^{d_1,\ldots,d_{\ell_j}}(\tau_1,\ldots,\tau_{\ell_j})
: = \left\{\begin{array}{cc}
             \Psi_j(\tau_1,\ldots,\tau_{\ell_j}), & k_1=\cdots=k_{\ell_j} = j, \ d_1=\cdots=d_{\ell_j}=-1, \\
             0, & \mathrm{otherwise},
           \end{array}
   \right.
\end{equation}
where the multi-variable function $\Psi_j(\tau_1,\ldots,\tau_{\ell_j})$ is defined in (\ref{eq:general_state_-1}). $\Psi_{k_1,\ldots,k_{\ell_j}}^{d_1,\ldots,d_{\ell_j}}(\tau_1,\ldots,\tau_{\ell_j})$ can be regarded as a $2\ell_j$-way tensor in the tensor space $\mathbb{C}^{\overbrace{m\times\ldots\times m}^{\ell_j} \times \overbrace{2 \times\ldots \times 2}^{\ell_j}}$. Accordingly, for each $j=1,\ldots, m$, and  $i = 1,\ldots, \ell_j$ define operators
\begin{equation*}\label{eq:b_d_2}
b_j^{d_i}(t)
:=
\left\{
\begin{array}{ll}
b_j^\ast(t), & d_i=-1, \\
b_j(t), & d_i=1.%
\end{array}%
\right.
\end{equation*}
Moreover, for each $j,k = 1,\ldots,m$, define
\begin{equation*}\label{eq:g_d}
g_{G^d}^{kj}(t)
:=
\left\{
\begin{array}{ll}
g_{G^-}^{kj}(t), & d=-1, \\
g_{G^+}^{kj}(t)^\ast, & d=1.%
\end{array}
\right.
\end{equation*}%
With the above notations, for each $j=1,\ldots,m$, $|\Psi_j\rangle$ defined in (\ref{eq:general_state_-1}) can be encoded by a $2\ell_j$-way tensor in the tensor space $\mathbb{C}^{\overbrace{m\times\ldots\times m}^{\ell_j} \times \overbrace{2 \times\ldots \times 2}^{\ell_j}}$. Specifically,
\begin{equation}\label{eq:psi_2}
|\Psi_j\rangle
=
\frac{1}{\sqrt{N_{\ell_j}}}\sum_{k_1,\ldots,k_{\ell_j}=1}^m \sum_{d_1,\ldots, d_{\ell_j}=\pm 1} (-1)^{\sum_{i=1}^{\ell_j}\mathrm{sgn}(d_i)}
\int_{\ell_j} \Psi_{k_1,\ldots,k_{\ell_j}}^{d_1,\ldots,d_{\ell_j}}(\tau_1,\ldots,\tau_{\ell_j}) b_j^{d_1}(\tau_1)\cdots b_j^{d_{\ell_j}}(\tau_{\ell_j})
d\tau_{1\to \ell_j}|0\rangle .
\end{equation}
Moreover, for each $j=1,\ldots,m$, $i = 1,\ldots, \ell_j$ and $k_i=1,\ldots,m$, define operators
\begin{equation} \label{eq:b_psi}
  b_{k_1,\ldots,k_{\ell_j}}^{d_1,\ldots,d_{\ell_j}}(\Psi_j) := \Psi_{k_1,\ldots,k_{\ell_j}}^{d_1,\ldots,d_{\ell_j}}(t_1,\ldots,t_{\ell_j}) b_j^{d_1}(t_1)\cdots b_j^{d_{\ell_j}}(t_{\ell_j}),
\end{equation}
where the $2\ell_j$-way tensor $\Psi_{k_1,\ldots,k_{\ell_j}}^{d_1,\ldots,d_{\ell_j}}(\tau_1,\ldots,\tau_{\ell_j})$ is that defined in (\ref{eq:psi_update_2}). Then (\ref{eq:psi_2}) becomes
\begin{equation*}
|\Psi_j\rangle
=
\frac{1}{\sqrt{N_{\ell_j}}}\sum_{k_1,\ldots,k_{\ell_j}=1}^m \sum_{d_1,\ldots, d_{\ell_j}=\pm 1} (-1)^{\sum_{i=1}^{\ell_j}\mathrm{sgn}(d_i)} \int_{\ell_j} b_{k_1,\ldots,k_{\ell_j}}^{d_1,\ldots,d_{\ell_j}}(\Psi_j)dt_1\ldots dt_{\ell_j}|0\rangle .
\end{equation*}
Accordingly, the multi-channel state $|\Psi\rangle$ in (\ref{eq:general_state_0}) can be re-written as
\begin{equation} \label{eq:psi_3}
|\Psi\rangle
=
\prod_{j=1}^m \frac{1}{\sqrt{N_{\ell_j}}}\sum_{k_1,\ldots,k_{\ell_j}=1}^m \sum_{d_1,\ldots, d_{\ell_j}=\pm 1} (-1)^{\sum_{i=1}^{\ell_j}\mathrm{sgn}(d_i)} \int_{\ell_j} b_{k_1,\ldots,k_{\ell_j}}^{d_1,\ldots,d_{\ell_j}}(\Psi_j)dt_1\ldots dt_{\ell_j}|0^{\otimes m}\rangle .
\end{equation}
The above motivates us to define a class of states.
\begin{defn} \label{def:F_2}
 Let $\Psi_{k_1,\ldots,k_{\ell_j}}^{d_1,\ldots,d_{\ell_j}}(\tau_1,\ldots,\tau_{\ell_j})$ be a $2\ell_j$-way tensor in the tensor space \newline $\mathbb{C}^{\overbrace{m\times\ldots\times m}^{\ell_j} \times \overbrace{2 \times\ldots \times 2}^{\ell_j}}$. A state $\rho_{\Psi,R}$ is said to be a
\emph{photon-Gaussian} state if it belongs to the set
\begin{align}
\mathcal{F}_2  & :=\left\{
\rho_{\Psi,R}
=
\prod_{j=1}^m \frac{1}{\sqrt{N_{\ell_j}}}\sum_{k_1,\ldots,k_{\ell_j}=1}^m \sum_{d_1,\ldots, d_{\ell_j}=\pm 1} (-1)^{\displaystyle{\sum_{i=1}^{\ell_j}\mathrm{sgn}(d_i)}} \int_{\ell_j} b_{k_1,\ldots,k_{\ell_j}}^{d_1,\ldots,d_{\ell_j}}(\Psi_j)dt_1\ldots dt_{\ell_j}|0^{\otimes m}\rangle \rho_R
\right. \nonumber \\
& \left. ~ \times \left(\prod_{j=1}^m \frac{1}{\sqrt{N_{\ell_j}}}\sum_{k_1,\ldots,k_{\ell_j}=1}^m \sum_{d_1,\ldots, d_{\ell_j}=\pm 1} (-1)^{\displaystyle{\sum_{i=1}^{\ell_j}\mathrm{sgn}(d_i)}} \int_{\ell_j} b_{k_1,\ldots,k_{\ell_j}}^{d_1,\ldots,d_{\ell_j}}(\Psi_j)dt_1\ldots dt_{\ell_j}|0^{\otimes m}\rangle \right)^\ast \right\} ,  \label{class_F}
\end{align}
where the operator $b_{k_1,\ldots,k_{\ell_j}}^{d_1,\ldots,d_{\ell_j}}(\Psi_j)$ is defined in (\ref{eq:b_psi}), and $\rho_R$ is a zero-mean Gaussian field state with covariance function $R$. It is assumed that  $\mathrm{Tr}[\rho_{\Psi,R}]=1$.
\end{defn}

\emph{Remark 7.} Clearly, the $m$-channel multi-photon state $|\Psi\rangle$ defined in (\ref{eq:general_state_0}) belongs to $\mathcal{F}_2$. Moreover, when $G$ is passive, $\mathcal{F}_1 = \mathcal{F}_2$.

Next we study how the input state in $\mathcal{F}_2$ is transformed by the quantum linear system $G$.

\begin{thm}\label{thm_active_2}
Suppose that the quantum linear system $G$ is asymptotically stable. The density function $\rho_{\Psi_{\rm out},R_{\rm out}}$ of the steady-state output field of $G$ driven by the density operator $\rho_{\Psi,R}\in\mathcal{F}_2$ is
\begin{eqnarray}
&&\rho_{\Psi_{\rm out},R_{\rm out}}
\label{eq:state_second}\\
&=&
\left(\prod_{j=1}^m\frac{1}{\sqrt{N_{\ell_j}}}\sum_{r_1,\ldots,r_{\ell_j}=1}^m\sum_{f_1,\ldots,f_{\ell_j}=\pm1}(-1)^{\sum_{i=1}^{\ell_j}\mathrm{sgn}(f_i)}
\int_{\ell_j}b_{r_1,\ldots,r_{\ell_j}}^{f_1,\ldots,f_{\ell_j}}(\Psi_{out,j}) d\tau_{1\to\ell_j}|0^{\otimes m}\rangle\right)\rho_{R_{\rm out}}
\nonumber\\
&& \ \
\times \left(\prod_{j=1}^m\frac{1}{\sqrt{N_{\ell_j}}}\sum_{r_1,\ldots,r_{\ell_j}=1}^m\sum_{f_1,\ldots,f_{\ell_j}=\pm1}(-1)^{\sum_{i=1}^{\ell_j}\mathrm{sgn}(f_i)}
\int_{\ell_j}b_{r_1,\ldots,r_{\ell_j}}^{f_1,\ldots,f_{\ell_j}}(\Psi_{out,j}) d\tau_{1\to\ell_j}|0^{\otimes m}\rangle\right),
\nonumber
\end{eqnarray}
where
\begin{equation*}
g_{G^{d_i,f_i}}^{kj}(t)
:=
\left\{
\begin{array}{ll}
g_{G^{-d_i}}^{kj}(t), & f_i=-1, \\
g_{G^{d_i}}^{kj}(t), & f_i=1,%
\end{array} \ \ \ \forall j,k = 1,\ldots,m, \ i=1,\ldots, \ell_j, \ d_i=\pm1 ,
\right.
\end{equation*}%
\begin{equation*} \label{eq:b_df}
b_i^{d_k,f_k}(t)
:=
\left\{
\begin{array}{ll}
b_i^{-d_k}(t), & f_i=-1, \\
b_i^{d_k}(t), & f_i=1,%
\end{array}%
\right.  \ \ \ \forall i=1,\ldots, m,  \ \ \ \forall k = 1,\ldots, \max\{\ell_1,\ldots,\ell_m\},
\end{equation*}%
\begin{align}
\Psi_{k_{1\to\ell_j},r_{1\to\ell_j}}^{d_{1\to\ell_j},f_{1\to\ell_j}}(t_1,\ldots,t_{\ell_j}) :=\int_{\ell_j}\prod_{i=1}^{\ell_j}g_{G^{d_i,f_i}}^{r_i k_i}(t_i-\tau_i)
\Psi_{k_1,\ldots,k_{\ell_j}}^{d_1,\ldots,d_{\ell_j}}(\tau_1,\ldots,\tau_{\ell_j})
d\tau_{1\to\ell_j} ,
\label{eq:temp8}
\end{align}
\begin{equation} \label{eq:b_d_3}
  b_{r_1,\ldots,r_{\ell_j}}^{f_1,\ldots,f_{\ell_j}}(\Psi_{out,j}) :=  \sum_{k_1,\ldots,k_{\ell_j}=1}^m
  \sum_{d_1,\ldots, d_{\ell_j}=\pm 1}  (-1)^{\sum_{i=1}^{\ell_j}\mathrm{sgn}(d_i)}\Psi_{k_{1\to\ell_j},r_{1\to\ell_j}}^{d_{1\to\ell_j},f_{1\to\ell_j}}(t_1,\ldots,t_{\ell_j})
  \prod_{i=1}^{\ell_j} b_{r_i}^{d_i,f_i}(t_i) ,
\end{equation}
and $\rho_{R_{\rm out}}$ is a Gaussian state whose covariance function is obtained by the Gaussian transfer
\begin{equation*}
R_{\rm out}[i\omega]=G[i\omega] R[i\omega]G[i\omega]^\dag.
\label{Phi_2}
\end{equation*}
\end{thm}

\textbf{Proof.}~
It is easy to show that (\ref{eq:b_minus}) can be re-written as
\begin{equation*}\label{eq:b_minus_2}
b_j^{-^d}(t,-\infty) = \sum_{k=1}^m \int_{-\infty}^\infty \left(-g_{G^{-d}}^{kj}(r-t)b_k^{-d}(r)+g_{G^d}^{kj}(r-t)b_k^d(r)\right)dr, \ \ d=\pm1 .
\end{equation*}
Note that
\begin{eqnarray*}
&&
\lim_{\begin{subarray}{c}t_0\rightarrow-\infty \\ t\rightarrow\infty\end{subarray}}U(t,t_0)\int_{\ell_j}\sum_{k_1,\ldots,k_{\ell_j}=1}^m\sum_{d_1,\ldots, d_{\ell_j}=\pm 1}(-1)^{\sum_{i=1}^{\ell_j}\mathrm{sgn}(d_i)}
\label{eq:output_state_general_2}\\
&&
\ \ \  \ \ \ \ \times \Psi_{k_{1\to\ell_j}}^{d_{1\to\ell_j}}(\tau_1,\ldots, \tau_{\ell_j})
b_{k_1}^{d_1}(\tau_1)\cdots b_{k_{\ell_j}}^{d_{\ell_j}}(\tau _{\ell_j})
 d\tau_{1\to\ell_j}U(t,t_0)^\ast
\nonumber\\
&=&
\int_{\ell_j}\sum_{k_1,\ldots,k_{\ell_j}=1}^m\sum_{d_1,\ldots, d_{\ell_j}=\pm 1}(-1)^{\sum_{i=1}^{\ell_j}\mathrm{sgn}(d_i)}
\Psi_{k_{1\to\ell_j}}^{d_{1\to\ell_j}}(\tau_1,\ldots, \tau_{\ell_j})
\nonumber\\
&&
\times \sum_{r_1=1}^m\int_{-\infty}^\infty\left(-g_{G^{-d_1}}^{r_1k_1}(t_1-\tau_1)b_{r_1}^{-d_1}(t_1)dt_1+g_{G^{d_1}}^{r_1k_1}(t_1-\tau_1)b_{r_1}^{d_1}(t_1)dt_1\right) \nonumber \\
&&
\cdots \sum_{r_{\ell_j}=1}^m\int_{-\infty}^\infty\left(-g_{G^{-d_{\ell_j}}}^{r_{\ell_j}k_{\ell_j}}(t_{\ell_j}-\tau_{\ell_j})b_{r_{\ell_j}}^{-d_{\ell_j}}(t_{\ell_j})dt_{\ell_j}+g_{G^{d_{\ell_j}}}^{r_{\ell_j}k_{\ell_j}}(t_{\ell_j}-\tau_{\ell_j})b_{r_{\ell_j}}^{d_{\ell_j}}(t_{\ell_j})dt_{\ell_j}\right)
d\tau_{1\to\ell_j}
\nonumber\\
&=&
\int_{\ell_j}\sum_{r_1,\ldots,r_{\ell_j}=1}^m\sum_{f_1,\ldots,f_{\ell_j}=\pm1}(-1)^{\sum_{i=1}^{\ell_j}\mathrm{sgn}(f_i)}
\sum_{k_1,\ldots,k_{\ell_j}=1}^m\sum_{d_1,\ldots, d_{\ell_j}=\pm 1}  (-1)^{\sum_{i=1}^{\ell_j}\mathrm{sgn}(d_i)}
\nonumber\\
&&
\ \ \ \times \Psi_{k_{1\to\ell_j},r_{1\to\ell_j}}^{d_{1\to\ell_j},f_{1\to\ell_j}}(t_1,\ldots,t_{\ell_j})\prod_{i=1}^{\ell_j} b_{r_i}^{d_i,f_i}(t_i)dt_{1\to\ell_j}
\nonumber\\
&=&
\sum_{r_1,\ldots,r_{\ell_j}=1}^m\sum_{f_1,\ldots,f_{\ell_j}=\pm1}(-1)^{\sum_{i=1}^{\ell_j}\mathrm{sgn}(f_i)}
\int_{\ell_j}b_{r_1,\ldots,r_{\ell_j}}^{f_1,\ldots,f_{\ell_j}}(\Psi_{\rm out}) dt_1\ldots dt_{\ell_j}
\end{eqnarray*}
This, together with the Gaussian transfer theorem (Theorem \ref{thm:spectral-transfer}), establishes Theorem \ref{thm_active_2}.

\emph{Remark 8.} It can be verified that the \emph{factorizable} $m$-channel multi-photon state $|\Psi\rangle$ defined in (\ref{Psi_in}) (equivalently (\ref{Psi_in_2})) can be re-written as%
\begin{equation} \label{Psi_in_2b}
|\Psi\rangle\langle\Psi|
=
\prod\limits_{j=1}^m \frac{1}{\sqrt{N_{\ell_j}}} \sum_{i=1}^m \prod\limits_{k=1}^{\ell_j}
 (B_j^\ast(\xi_{ijk}^-)-B_j(\xi_{ijk}^+))|0^{\otimes m}\rangle \langle 0^{\otimes m}|\left(\prod\limits_{j=1}^m \frac{1}{\sqrt{N_{\ell_j}}} \sum_{i=1}^m \prod\limits_{k=1}^{\ell_j}
 (B_j^\ast(\xi_{ijk}^-)-B_j(\xi_{ijk}^+))\right)^\ast  .
\end{equation}
There is clear similarity between (\ref{Psi_in}) and (\ref{eq:psi_3}), or equivalently, between (\ref{Psi_in_2b}) and (\ref{class_F}). The implication of such similarity is that all the results for the unfactorizable case can be reduced to those for the factorizable case.

\emph{Remark 9.}  When the quantum linear system $G$ is passive and $\rho_R=|\phi\rangle \langle\phi|$, $\rho_{\Psi,R}$  in (\ref{class_F}) becomes a pure state. Moreover, for the case case, ${\rm sgn}(d_i)=0$  for all $i$. Therefore, in the passive case  $\rho_{\Psi,R}$ is a pure state in the class $\mathcal{F}_1$ defined in (\ref{eq:F_1}). As a result, in the passive case Theorem \ref{thm_active_2} reduces to Theorem \ref{thm:passive_multi_channel}.

\emph{Remark 10.}  From (\ref{eq:temp8}) it can be seen that $\Psi_{k_{1\to\ell_j},r_{1\to\ell_j}}^{d_{1\to\ell_j},f_{1\to\ell_j}}(t_1,\ldots,t_{\ell_j})$ is a $4\ell_j$ way tensor, not a $2\ell_j$ way tensor in the space $\mathbb{C}^{\overbrace{m\times\ldots\times m}^{\ell_j} \times \overbrace{2 \times\ldots \times 2}^{\ell_j}}$. As a result, $\rho_{\Psi_{\rm out},R_{\rm out}}$ in  (\ref{eq:state_second}) is not an element in the class $\mathcal{F}_2$. That is, the class $\mathcal{F}_2$ is not an invariant set under the steady-state action of the quantum linear system $G$. However, using a procedure similar to that presented in Theorem \ref{thm_active_2}, it is not hard to derive the steady-state output state when a quantum linear system $G$ is driven by an input state $\rho_{\Psi_{\rm out},R_{\rm out}}$. Clearly, the tensor representation plays a key role in this study.

Next we use three examples to illustrate the results for the unfactorizable photon states.

\emph{Example 2: The $(1+\ell)$-photon case.} Consider a beamsplitter with parameter
\[
S
=
\left[
\begin{array}{cc}%
\sqrt{1-R} & \sqrt{R}\\
\sqrt{R} & -\sqrt{1-R}%
\end{array}
\right]  , \ \ \ (0<R<1).
\]
Let the input state be%
\[
|\Psi_{in}\rangle
=
B_1^\ast(\xi)\otimes\frac{1}{\sqrt{N_\ell} }\prod\limits_{k=1}^\ell B_2^\ast(\xi_k)|00\rangle.
\]
As with Example 1, the output state can be derived by means of Corollary \ref{cor:output_state_multiple_channel_passive}. Alternatively, it can be derived via Theorem \ref{thm:passive_multi_channel}. Clearly, $m=2$, $\ell_1=1$, and $\ell_2=\ell$. By Theorem \ref{thm:passive_multi_channel}, the output state is
\begin{eqnarray}
 |\Psi_{\rm out}\rangle
  =
   (\sqrt{1-R}B_1^\ast(\xi)+\sqrt{R}B_2^\ast(\xi)) \frac{1}{\sqrt{N_\ell}}\sum_{k_1,\ldots,k_\ell=1}^2 B_{k_1}^\ast(S^{k_12}\xi_{k_1})\cdots B_{k_\ell}^\ast(S^{k_\ell2}\xi_{k_\ell})|00\rangle. \label{ex:bm:out}
\end{eqnarray}
In particular, assume
$\xi_1(t)\equiv\cdots\equiv\xi_\ell(t)\equiv\xi(t)$
and
$\int_{-\infty}^\infty|\xi(t)\vert^2dt=1$. Then (\ref{ex:bm:out}) becomes%
\[
|\Psi_{\rm out}\rangle
=
\frac{1}{\sqrt{\ell!}}(\sqrt{1-R}B_1^\ast(\xi)+\sqrt{R}B_2^\ast(\xi))
(\sqrt{R}B_1^\ast(\xi)-\sqrt{1-R}B_2^\ast(\xi))^\ell |00\rangle.
\]
The coefficient for the component
$\frac{1}{\sqrt{\ell!}}B_1^\ast(\xi)^\ell B_2^\ast(\xi)|00\rangle=\frac{1}{\sqrt{\ell!}}|\ell,1\rangle $
is $\sqrt{R^{\ell-1}}(R-\ell(1-R))$,
whose squared value is exactly (in) in \cite{SRZ06}.

\emph{Example 3: The photon-catalyzed optical coherent (PCOC) case.} Consider a beamsplitter with parameter
\begin{equation*}\label{S_supp}
  S = \left[ \begin{array}{cc}
                 T & -R \\
                 R & T
               \end{array}
    \right], \ \ \ (R,T>0, ~~R^2+T^2=1).
\end{equation*}
Let the input be $|\psi_\ell\rangle\otimes|\alpha\rangle$, where $|\alpha\rangle = e^{-|\alpha|^2/2}\sum_{n=0}^\infty \frac{\alpha^n}{\sqrt{n!}}|n\rangle$ is a coherent state.
The input stat can be re-written as
\begin{equation*}\label{in_supp}
|\Psi_{in}\rangle
=
e^{-|\alpha|^2/2}\sum_{n=0}^\infty\frac{\alpha^n}{\sqrt{n!}} |\ell\rangle\otimes|n\rangle =
e^{-|\alpha|^2/2}\sum_{n=0}^\infty\frac{\alpha^n}{\sqrt{n!}} \prod_{j=1}^2\prod_{k=1}^{\ell_j} B_j^\ast(\xi)|0^{\otimes2}\rangle,
\end{equation*}
where $\ell_1=\ell$ and $\ell_2=n$. That is, here it is assumed that all the photons are identical. 
By (\ref{eta}),
\begin{eqnarray*} \label{eta_supp}
\eta_{11k}^- = T\xi,  \ \  \eta_{12k}^- = -R\xi, \ \ \   \eta_{21r}^- = R\xi,  \ \   \eta_{22r}^- = T\xi, \ \ \forall   k=1,\ldots,\ell, \ \ \ \forall r=1,\ldots,n .
\end{eqnarray*}
By Theorem \ref{thm:passive_multi_channel},
\begin{eqnarray*}
|\Psi_{\rm out}\rangle &=& e^{-|\alpha|^2/2}\sum_{n=0}^\infty\frac{\alpha^n}{\sqrt{n!}} \prod_{j=1}^2 \prod_{k=1}^{\ell_j}\sum_{i=1}^2B_i^\ast(\eta_{ijk}^-)|0^{\otimes2}\rangle  \label{Psi_out_supp}
\\
&=&
e^{-|\alpha|^2/2}\sum_{n=0}^\infty\frac{\alpha^n}{\sqrt{n!}} \sum_{i=0}^\ell \sum_{j=0}^n \binom{n}{n-j}\binom{\ell}{i}(-1)^j T^{n+\ell-i-j} R^{i+j}|\ell+j-i\rangle\otimes|n+i-j\rangle.
\nonumber
\end{eqnarray*}
When the first output channel is measured by means of the state $|\ell\rangle$, the state at the second output channel becomes
\begin{equation*} \label{Psi_out2_supp}
|\Psi_{\rm out,conditioned}\rangle = e^{-|\alpha|^2/2}\sum_{n=0}^\infty\frac{\alpha^n}{\sqrt{n!}} \sum_{j=0}^{\min\{\ell,n\}} \binom{n}{n-j}\binom{\ell}{j}(-1)^j T^{n+\ell-2j} R^{2j}|n\rangle,
\end{equation*}
which reproduces the key formula (1) in \cite{BDS+12}.

 \emph{Remark 11.} Examples 1, 2, and 3 illustrate that the proposed research is also applicable to the discrete-variable multi-photon case.

\begin{figure}[ptb]
\centering
\includegraphics[width=6.2in]{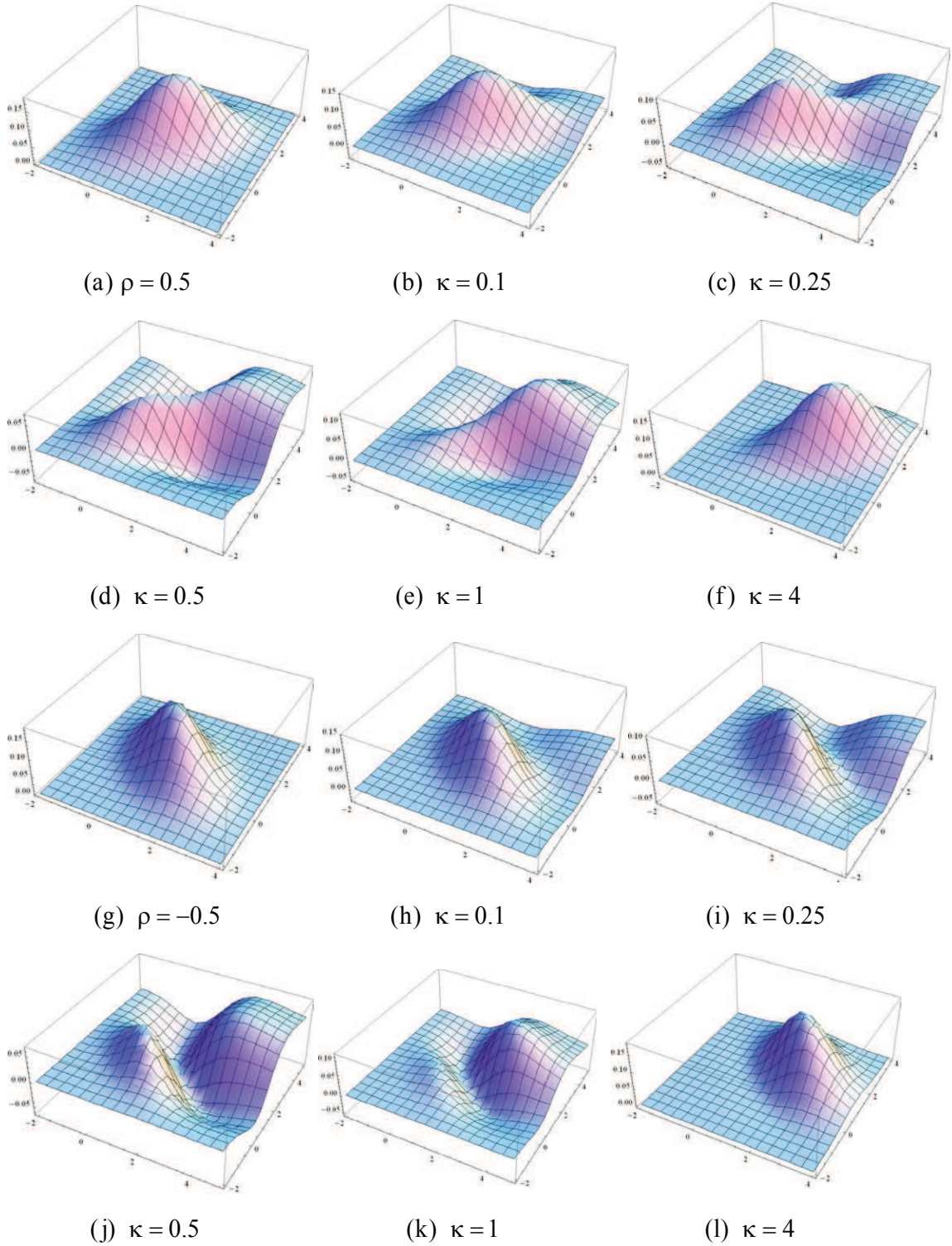}\caption{Multi-photon pulse shaping via an optical cavity. (a) and (g) are pulse shapes for the input 2-photon state with different correlations $\rho$. When the decay rate is small ((b) and (h)), the shape of $\psi_{\rm out}^-$ of the output 2-photon state is similar to that of $\psi$ of the input 2-photon state; As decay rate increases, the pulse shapes deform ((c)-(e) and (i)-(k)); When the decay rate is large ((f) and (l)), the shape of $\psi_{\rm out}^-$ is similar with that of $\psi$, however their mean values are significantly different.}
\label{fig_example_4}%
\end{figure}

\emph{Example 4: Multi-photon pulse shaping via optical cavity.}  In this example, we study how an optical cavity responds to an unfactorizable 2-photon input state. The optical cavity has parameters
$\Omega_- = \Omega_+ =0, C_- = \sqrt{\kappa}, C_+ = 0, S = 1$.
Thus,
$A=-\frac{\kappa}{2}I_2, B=-\sqrt{\kappa}I_2, C=\sqrt{\kappa}I_2, D=I_2$.
Let an unfactorizable 2-photon input state be %
\[
|\psi_2\rangle
=
\frac{1}{\sqrt{N_2}} \int_{-\infty}^\infty \int_{-\infty}^\infty \psi(t_1,t_2) b^\ast(t_1) b^\ast(t_2)dt_1 dt_2 |0\rangle,
\]
 where
\begin{equation*}
\psi(t_1,t_2)
=
\frac{1}{2\pi \left\vert \Sigma \right\vert^{1/2}}
\exp \left( -\frac{1}{2}\left[
\begin{array}{cc}
t_1-\tau _1 & t_2-\tau_2%
\end{array}%
\right] \Sigma^{-1}\left[
\begin{array}{c}
t_1-\tau_1 \\
t_2-\tau_2%
\end{array}%
\right] \right),
\label{example_xi}
\end{equation*}
with
\begin{equation*}
\Sigma =\left[
\begin{array}{cc}
\sigma_1^{2} & \rho \sigma_1\sigma_2 \\
\rho \sigma_1\sigma_2 & \sigma_2^2%
\end{array}%
\right], \ \ \ \sigma_1>0, \ \sigma_2>0, \  -1< \rho < 1.
\end{equation*}%
That is, the input state has a 2-dimensional Gaussian pulse shape centered at $(\tau_1,\tau_2)$ and with covariance matrix $\Sigma$. When the correlation parameter $\rho=0$, $|\psi_2\rangle$ reduces to a factorizable state.  According to Corollary \ref{thm:general_state}, the steady-state output state $|\psi_{\rm out}^-\rangle$ is given by
\begin{equation*}
\psi_{\rm out}^-(t_1,t_2)
=
\int_{-\infty}^\infty \int_{-\infty}^\infty g_{G^-}(t_1-r_1)g_{G^-}(t_2-r_2)\psi (
r_1,r_2) dr_1 dr_2,
\end{equation*}
where
\[
g_{G^-}(t)=
\left\{
\begin{array}{cc}%
\delta(t)-\kappa e^{-\frac{\kappa}{2}t}, & t\geq0,\\
0, & t<0.
\end{array}
\right.
\]
In the following we fix $\tau_1 = \tau_2 = \sigma_1 = \sigma_ 2 =1$, and study the pulse shape $\psi_{\rm out}^-(t_1,t_2)$ of the output state for several pairs of the correlation coefficient $\rho$ and the cavity decay rate $\kappa$. Fig.~\ref{fig_example_4} summarizes pulse shaping of multi-photon states by the optical cavity in different scenarios. Fig. 2(a)-(f) are for the case of $\rho=0.5$. Fig.~2(a) is the shape $\psi(t_1,t_2)$ of the input state, while  Fig.~2(b)-(f) are the shapes $\psi_{\rm out}^-(t_1,t_2)$ for the output state for different decaying rates.    Fig.~2(g)-(k) are for the case of $\rho=-0.5$. Fig.~2(g) is the shape $\psi(t_1,t_2)$ of the input state, while  Fig.~2(h)-(k) are the shapes $\psi_{\rm out}^-(t_1,t_2)$ for the output state for different decaying rates.

\section{Conclusion}\label{sec:conclusion}
In this paper we have studied the response of quantum linear systems to multi-channel multi-photon states. New types of tensors are defined to encode pulse information of multi-photon states, for both the factorizable case and the unfactorizable case. The steady-state action of quantum linear systems on multi-photon states are characterized in terms of tensor processing by transfer functions.  Explicit forms of output states, output covariance functions and output intensities have been derived. In contrast to the discrete-variable (single-mode) treatments in most discussions on quantum information, we have presented a continuous-variable (multi-mode) treatment of multi-photon processing. As can be seen from Examples  1-3, the continuous-variable treatment is also applicable to many discrete-variable treatments. Moreover, the continuous-variable treatment is closer to a real experimental environment in optical quantum information processing. As demonstrated by Example 4 for pulse shaping by optical cavities, one immediate future research is: How to design desired pulse shapes (which encode time or frequency correlation among photons) by means of quantum linear systems, as has been investigated in \cite{Milburn08} and \cite{ZJ13} in the single-photon setting for the passive case. Another future research is to study how multi-photon pulses can be stored and read out by gradient echo memories (\cite{HCH+13}), which are indispensable components of complex quantum optical networks for quantum communication and computing.

\end{document}